\documentclass[sigplan,10pt]{acmart}

\AtBeginDocument{%
  }

\setcopyright{acmcopyright}
\copyrightyear{2023} 
\acmYear{2023} 
\setcopyright{acmlicensed}\acmConference[SOSP '23]{ACM SIGOPS 29th Symposium on Operating Systems Principles}{October 23--26, 2023}{Koblenz, Germany}
\acmBooktitle{ACM SIGOPS 29th Symposium on Operating Systems Principles (SOSP '23), October 23--26, 2023, Koblenz, Germany}
\acmPrice{15.00}
\acmDOI{10.1145/3600006.3613144}
\acmISBN{979-8-4007-0229-7/23/10}

\usepackage{tikz}
\usepackage{amsmath}
\usepackage{xspace}
\usepackage{enumitem}
\usepackage{booktabs}
\usepackage{bbding}
\usepackage{subfig}
\usepackage{listings}
\usepackage[font=small]{caption}
\newcommand{\dmc}{Ditto\xspace}

\newcommand{\ie}{{\em i.e.},\xspace}
\newcommand{\eg}{{\em e.g.},\xspace}

\newcommand{\afflia}[0]{$^{\dag}$}
\newcommand{\afflib}[0]{$^{\mathsection}$}
\newcommand{\afflic}[0]{$^{\ddag}$}
\newcommand{\afflid}[0]{$^{\mathparagraph}$}

\makeatletter
\newcommand\footnoteref[1]{\protected@xdef\@thefnmark{\ref{#1}}\@footnotemark}
\makeatother

\begin{document}

\date{}

\title{\dmc: An Elastic and Adaptive Memory-Disaggregated Caching System}


\author{
    {\rm Jiacheng Shen\afflia\textsuperscript{1}, }
    {\rm Pengfei Zuo\afflib, }
    {\rm Xuchuan Luo\afflic\textsuperscript{1}, }
    {\rm Yuxin Su\afflid, }\\
    {\rm Jiazhen Gu\afflia, }
    {\rm Hao Feng\afflib, }
    {\rm Yangfan Zhou\afflic, }
    {\rm Michael R. Lyu\afflia}
}

\thanks{\textsuperscript{1} Work mainly done during the internship at Huawei Cloud.}

\author{
    \afflia The Chinese University of Hong Kong, 
    \afflib Huawei Cloud,
    \afflic Fudan University,
    \afflid Sun Yat-sen University
}










\acmArticleType{Research}

\begin{abstract}
    \noindent
    In-memory caching systems are fundamental building blocks in cloud services.
    However, due to the coupled CPU and memory on monolithic servers, existing caching systems cannot elastically adjust resources in a resource-efficient and agile manner.
    To achieve better elasticity, we propose to port in-memory caching systems to the disaggregated memory (DM) architecture, where compute and memory resources are decoupled and can be allocated flexibly.
    However, constructing an elastic caching system on DM is challenging since accessing cached objects with CPU-bypass remote memory accesses hinders the execution of caching algorithms.
    Moreover, the elastic changes of compute and memory resources on DM affect the access patterns of cached data, compromising the hit rates of caching algorithms.
    We design \dmc, the first caching system on DM, to address these challenges.
    \dmc first proposes a \textit{client-centric caching framework} to efficiently execute various caching algorithms in the compute pool of DM, relying only on remote memory accesses.
    Then, \dmc employs a \textit{distributed adaptive caching scheme} that adaptively switches to the best-fit caching algorithm in real-time based on the performance of multiple caching algorithms to improve cache hit rates.
    Our experiments show that \dmc effectively adapts to the changing resources on DM and outperforms the state-of-the-art caching systems by up to $3.6\times$ in real-world workloads and $9\times$ in YCSB benchmarks.
\end{abstract}



\settopmatter{printfolios=false}
\settopmatter{printacmref=false}
\maketitle
\pagestyle{empty}

\section{Introduction}\label{sec:intro}
\noindent
In-memory caching systems, \eg Memcached~\cite{memcached} and Redis~\cite{redis}, are widely adopted in cloud services~\cite{hotmetrics12atikoglu,nsdi21yang,ecommerse-cache,gaming-cache} to reduce service latency and improve throughput.
Due to the dynamic and bursty characteristics of requests in cloud services~\cite{atc20shahrad,nsdi22weng,nsdi15suresh}, elasticity, \ie the ability to adjust compute and memory resources according to workload changes, is a critical requirement for in-memory caching systems.

However, existing caching systems are constructed with and deployed on monolithic servers with coupled CPU and memory, which has two issues in dynamic resource adjustments.
First, resource utilization is compromised since CPU and memory have to be added or reduced \textit{together} as fix-sized virtual machines (VMs) on monolithic servers~\cite{elasticache,eurosys18nitu}. 
While in practice, services may only want to add more memory or CPU cores to increase either cache capacity or request throughput.
Besides, the speed of adjusting resources is too slow to cope with the workload bursts due to the time-consuming data migration~\cite{sosp17kulkarni,sigmod11elmore}.

Disaggregated memory (DM)~\cite{osdi18shan,isca09lim,asplos22guo,sosp21lee,osr23Aguilera} is a promising approach to address these issues.
It decouples the CPU and memory of monolithic servers into independent compute and memory pools and connects them with high-speed CPU-bypass interconnects, \eg remote direct memory access (RDMA)~\cite{infiniband} and compute express link (CXL)~\cite{cxl}.
CPUs and memory can thus be independently adjusted as application demands, improving resource efficiency.
Moreover, the frequency of data migration can be greatly reduced since data are shared by all CPU cores in the compute pool and only need to be migrated to achieve better load balancing~\cite{vldb22lee,fast23shen}.
However, two challenges have to be addressed to achieve a practical caching system on DM.

\textit{1) Bypassing remote CPUs hinders the execution of caching algorithms.}
Caching systems use various caching algorithms under different workloads~\cite{redis,atc17blankstein}.
Caching algorithms monitor the hotness of cached objects and select eviction victims by maintaining the hotness information in caching data structures.
Since data access changes object hotness, existing caching algorithms rely on the CPUs of caching servers, where all data accesses are executed, to monitor object hotness and maintain caching data structures~\cite{memcached}.
However, in caching systems on DM, applications (clients) in the compute pool bypass CPUs in the memory pool when accessing objects.
Evaluating object hotness becomes difficult due to the lack of a centralized hotness monitor on data paths.
Selecting eviction victims becomes inefficient since caching data structures have to be maintained with multiple high-latency remote memory accesses by clients, where data accesses are executed.
Moreover, supporting various caching algorithms for different workloads~\cite{redis,atc17blankstein} is even more difficult on DM since caching algorithms evict objects with specified rules and tailored data structures~\cite{2q,hlru}.

\textit{2) Adjusting resources affects hit rates of caching algorithms.}
Hit rates of caching algorithms relate to the data access patterns of workloads~\cite{hotstorage18vietri} and the cache size~\cite{fast21rodriguez}. 
On DM, both attributes change on dynamical resource adjustments.
The data access pattern changes with the number of concurrent clients (\ie compute resources), and the cache size changes with the allocated memory spaces (\ie memory resources).
As a result, the best caching algorithm that maximizes hit rate changes dynamically with resource settings.
Caching systems with fixed caching algorithms cannot adapt to these dynamic features of DM and can lead to inferior hit rates.

We address these challenges with \dmc\footnote{\dmc is a Pok\'{e}mon that can arbitrarily change its appearance.}, an elastic and adaptive caching system on DM.
First, we propose a client-centric caching framework with \textit{distributed hotness monitoring} and \textit{sample-based eviction} to address the challenges of executing caching algorithms on DM.
The distributed hotness monitoring uses one-sided RDMA verbs to record the access information from distributed clients in the compute pool, uses eviction priority to formally describe object hotness, and assesses objects' eviction priorities by applying priority calculation rules on the recorded access information.
The sample-based eviction scheme selects eviction victims by sampling multiple objects and selecting the one with the lowest priority on the client side without maintaining remote data structures~\cite{redis}.
Since the key difference among caching algorithms is their definitions of eviction priorities, various caching algorithms can be integrated by defining tailored priority calculation rules with little coding effort.
Second, we propose a distributed adaptive caching scheme to address the challenge of dynamic resource change.
\dmc simultaneously executes multiple caching algorithms with the client-centric caching framework and uses regret minimization~\cite{soda05flaxman,corr15foster,yusuf2020cache}, an online machine learning algorithm, to perceive their performance and select the best one in the current resource setting.

We implement \dmc and evaluate its performance with both synthesized and real-world workloads~\cite{fast10koller,nsdi20song,osdi20yang}.
\dmc is more elastic than Redis regarding resource efficiency and the speed of resource adjustments.
On YCSB and real-world workloads, \dmc outperforms CliqueMap~\cite{sigcomm21singhvi}, the state-of-the-art key-value cache, by up to $9\times$ and $3.6\times$, respectively.
Moreover, \dmc can flexibly extend 12 widely-used caching algorithms with $12.5$ lines of code (LOC) on average. 
The implementation of \dmc is open-source\footnote{\url{https://github.com/dmemsys/Ditto.git}.}.

The contributions of this paper include the following:
\begin{itemize}
    \item We identify the elasticity benefits and challenges of constructing caching systems on DM and propose \dmc, the first caching system on DM.
    \item We propose a client-centric caching framework where various caching algorithms can be integrated flexibly and executed efficiently on DM. A sample-friendly hash table and a frequency counter cache are designed to improve the efficiency of the framework on DM.
    \item We propose distributed adaptive caching to provide high hit rates by selecting the best caching algorithm according to the dynamic resource change and various data access patterns on DM. A lightweight eviction history and a lazy weight update scheme are designed to efficiently achieve adaptivity on DM.
    \item We implement \dmc and evaluate it with various workloads. \dmc outperforms the state-of-the-art approaches by up to $9\times$ under YCSB synthetic workloads and up to $3.6\times$ under real-world workloads.
\end{itemize}
\section{Background and Motivation}\label{sec:background}

\subsection{Issues of Caching Systems on Monolithic Servers}
\noindent 
There are two issues with existing caching systems on monolithic servers when they adjust resources.

\textit{1) Resource inefficiency.}
Resources of existing caching services on monolithic servers, \eg ElastiCache~\cite{elasticache}, are allocated with fix-sized virtual machines (VMs) with both CPU and memory, \eg 1 CPU with 2 GB DRAM, to facilitate resource management in monolithic-server-based datacenters.
Resources are wasted when coupled CPU and memory are allocated, but only CPU or memory needs to be dynamically increased. 
Moreover, applications' demands on resources must be rounded up to fit in these fix-sized VMs, causing low resource utilization in the entire datacenter~\cite{techreport2009armbrust}.

\begin{figure}
    \centering
    \includegraphics[width=0.95\columnwidth]{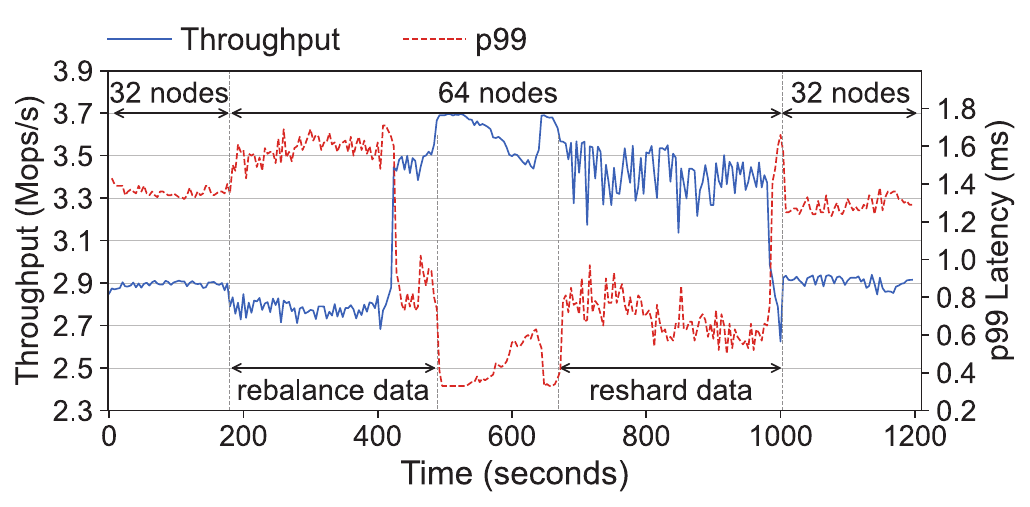}
    \caption{The performance of Redis when adjusting resources.}\label{fig:redis-elasticity-issue}
    \vspace{-0.1in}
\end{figure}

\textit{2) Slow resource adjustments.}
Existing in-memory caching systems shard data to multiple VMs to leverage more CPU and memory resources~\cite{nsdi14lim,elasticache,memorystore,redis}.
Cached data have to be resharded and migrated when new VMs are added to the caching cluster.
The migration cost~\cite{sigmod11elmore} is unavoidable when either CPU or memory needs to be adjusted due to the coupled allocation of CPU and memory on monolithic servers.
The performance gain when increasing resources and the resource reclamation after shrinking resources is delayed for minutes due to the time-consuming data migration~\cite{elasticachedd}.
Moreover, the throughput drops and latency increases due to the consumption of additional CPU cycles and network bandwidths spent on moving data~\cite{sosp17kulkarni,cikm12qin}.

Figure~\ref{fig:redis-elasticity-issue} shows the migration cost on Redis~\cite{redis}, the backend of many cloud caching services~\cite{elasticache,memorystore}, during resource adjustments under the read-only YCSB-C workload~\cite{socc10cooper} with 10 million 256B key-value pairs.
We first use 32 Redis nodes, each with 1 CPU core and 1 GB DRAM, then add 32 more nodes after 3 minutes of execution, and shrink back to 32 nodes after 3 minutes of stable execution with 64 nodes.
We launch all 64 Redis nodes and idle 32 of them initially to rule out the cost of starting Redis nodes.
We use 512 client threads to get the maximum throughput.
When scaling to 64 nodes, Redis takes 5.3 minutes to migrate data.
The throughput drops up to 7\%, and the 99th percentile latency increases up to 21\% in the process.
When shrinking back to 32 nodes, the resource reclamation is delayed for 5.6 minutes due to data migration.
Such migration cost is unavoidable even if using advanced migration techniques~\cite{sosp17kulkarni,sigmod11elmore} since CPU and memory are allocated in a coupled manner in VMs and objects are sharded to individual VMs. 

\subsection{Disaggregated Memory}
\noindent
Disaggregated memory (DM) is proposed to reduce the total cost of ownership (TCO) and improve the elasticity of applications on cloud datacenters~\cite{osdi18shan,atc20shahrad,osdi20wang}.
It decouples compute and memory resources of monolithic servers into autonomous compute and memory pools.
The compute pool contains compute nodes (CNs) with abundant CPU cores and a small amount of DRAM serving as run-time caches.
The memory pool holds memory nodes (MNs) with adequate memory and a controller with weak compute power (\eg 1 - 2 CPU cores) to execute management tasks, \ie network connection and memory management.
CNs and MNs are connected with CPU-bypass interconnects with high bandwidth and microsecond-scale latency, \eg RDMA and CXL~\cite{cxl}, ensuring the performance requirements of memory accesses.
CNs can allocate and free variable-sized memory blocks in the memory pool through the \textsf{ALLOC} and \textsf{FREE} interfaces provided by the controller.
Without loss of generality, in this paper, we assume that CNs access MNs through one-sided RDMA verbs, \ie \textsf{READ}, \textsf{WRITE}, \textsf{ATOMIC\_CAS} (compare and swap), and \textsf{ATOMIC\_FAA} (fetch and add).

The decoupled compute and memory resources of DM addresses the resource efficiency and elasticity issues of existing caching systems.
First, with DM, compute and memory resources can be allocated separately in a fine-grained manner~\cite{osdi20wang}.
Resources can be used more efficiently by assigning the exact amount of resources as per application demands.
Second, the frequency of data migration can be greatly reduced.
Specifically, caching systems on DM do not need to migrate data when expanding or reducing memory since the cached data in the memory pool can be accessed by all CNs in the compute pool.
Only in some special cases, \eg the network bandwidth of an MN becomes the performance bottleneck due to skewed workloads, data migration happens to achieve better load balancing.
As a result, the migration cost can be eliminated for most cases, allowing resource adjustments to take effect agilely without performance losses.
\section{Challenges}\label{sec:challenges}

\subsection{Executing Caching Algorithms on DM}
\noindent
Existing caching algorithms are designed for \textit{server-centric} caching systems on monolithic servers where all data are accessed and evicted by the server-side CPUs in a centralized manner.
Such a setting, however, no longer holds on DM because 1) caching systems on DM are \textit{client-centric}, where clients directly access and evict the cached data in a CPU-bypass manner, and 2) the compute power in the memory pool of DM is too weak to execute caching algorithms on the data path.
Two problems need to be addressed to execute caching algorithms on DM.

The first problem is how to evaluate the hotness of cached objects in the \textit{client-centric} setting.
Existing caching algorithms assess objects' hotness by monitoring and counting all data accesses~\cite{hlru,gdsf,gds}.
The monitoring can be trivially achieved on server-centric caching systems since the CPUs of monolithic caching servers access all data.
However, on DM, accesses to cached objects cannot be monitored either in the memory pool or on clients because 1) RDMA bypasses the CPUs in the memory pool, and 2) individual clients in the compute pool are not aware of global data accesses.

\begin{figure}[t]
    \vspace{-0.15in}
    \centering
    \subfloat[Single-client performance.]{
        \includegraphics[width=0.5\columnwidth]{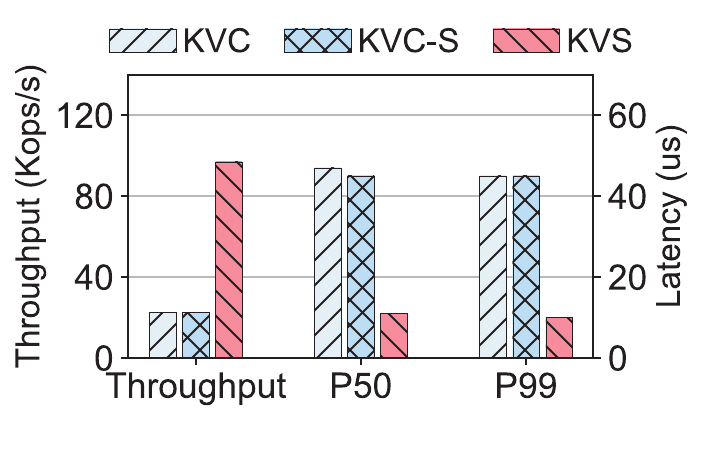}
        \label{fig:plru-1}
    }
    \hspace{1mm}
    \subfloat[Multi-client throughput.]{
    \includegraphics[width=0.43\columnwidth]{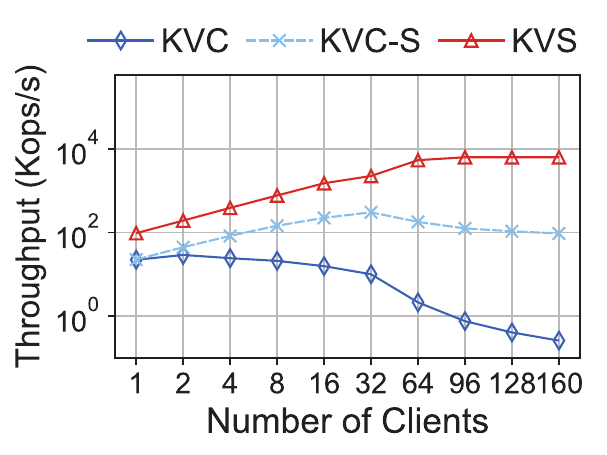}
        \label{fig:plru-2}
    }
    \caption{The cost of maintaining caching data structures on DM.}\label{fig:bg-cost}
    \vspace{-0.1in}
\end{figure}

The second problem is how to efficiently select eviction victims on the client side.
Caching algorithms maintain various caching data structures, \eg lists~\cite{hlru}, heaps~\cite{gdsf,gds}, and stacks~\cite{lirs}, to reflect the hotness of cached objects and select eviction victims based on these data structures.
The data structures are maintained by the CPUs of caching servers on each data access since access changes object hotness.
However, the maintenance of caching data structures has to be executed by clients in the compute pool since clients directly access objects with one-sided RDMA verbs.
Maintaining these data structures thus becomes inefficient due to the multiple RTTs required on the critical path.
Besides, locks are required to ensure the correctness of caching data structures under concurrent accesses~\cite{memcached}.
The throughput of caching systems will be severely bottlenecked by the microsecond-scale lock latency and the network contention caused by iteratively retying on lock failures~\cite{sigmod22wang}.

To show the problem of maintaining caching data structures, we compare the performance of a linked-list-based LRU key-value cache (KVC), a key-value cache with sharded LRU lists (KVC-S), and a key-value store (KVS) on DM~\cite{fast23shen} under the read-only YCSB-C benchmark~\cite{socc10cooper}.
All approaches use a lock-free hash table to index cached objects.
KVC maintains a lock-protected linked list to execute LRU.
KVC-S shards the LRU list into 32 sub-lists to avoid lock contention and sleeps 5 us on lock failures to reduce the wasted RDMA requests on lock failures.
Figure~\ref{fig:plru-1} shows the throughput and latency of the three approaches with a single client, ruling out lock contention.
The throughput of KVC and KVC-S is only $23\%$ of that of KVS, and the tail latency is more than $4.5\times$ higher due to the additional RDMA operations on the critical path of data accesses.
Figure~\ref{fig:plru-2} shows their throughput with growing numbers of client threads.
The throughputs of KVC and KVC-S drop with more than 32 client threads because the RNIC of the MN is overwhelmed by the useless \textsf{RDMA\_CAS}es on lock-fail retries.
The throughput of KVC-S drops more mildly due to the 5 us backoff on lock failures.

\subsection{Dynamic Resource Changes Affect Hit Rate}
\noindent
Hit rates of caching algorithms closely relate to the data access patterns and the cache size~\cite{fast21rodriguez}.
However, both aspects are affected when dynamically adjusting compute and memory resources, making the best caching algorithm that maximizes the hit rate changes accordingly.
Since DM enables resources to be adjusted fleetly and frequently, the effect of changing resource settings is amplified.
Caching systems with fixed caching algorithms cannot adapt to these dynamic features of DM and can lead to inferior hit rates.

\begin{figure}
    \begin{minipage}[t]{0.45\columnwidth}
        \centering
        \includegraphics[width=\columnwidth]{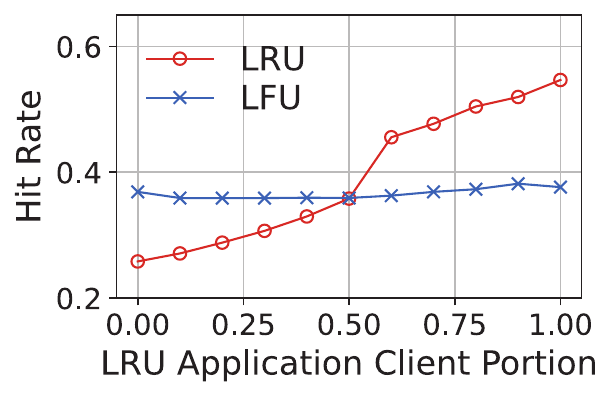}
        \caption{Hit rates under different numbers of clients under different applications.}
        \label{fig:bg1}
    \end{minipage}%
    \hspace{5mm}
    \begin{minipage}[t]{0.45\columnwidth}
        \centering
        \includegraphics[width=\columnwidth]{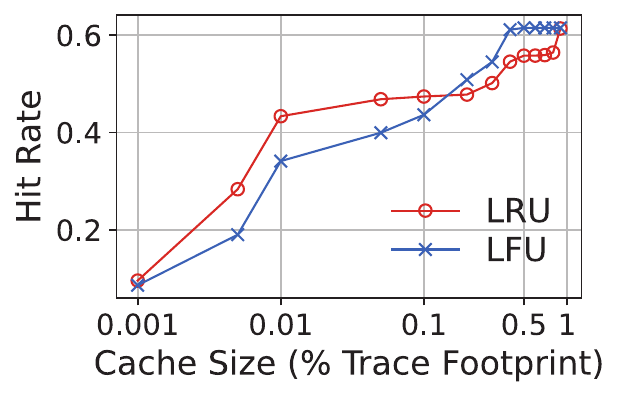}
        \caption{Hit rates of LRU and LFU on the same workload with different cache sizes.}
        \label{fig:bg4}
    \end{minipage}
    \vspace{-0.1in}
\end{figure}

\begin{figure}[t]
    \vspace{-0.15in}
    \centering
    \subfloat[The CDF of relative hit rate changes on 74 workloads.]{
        \includegraphics[width=0.45\columnwidth]{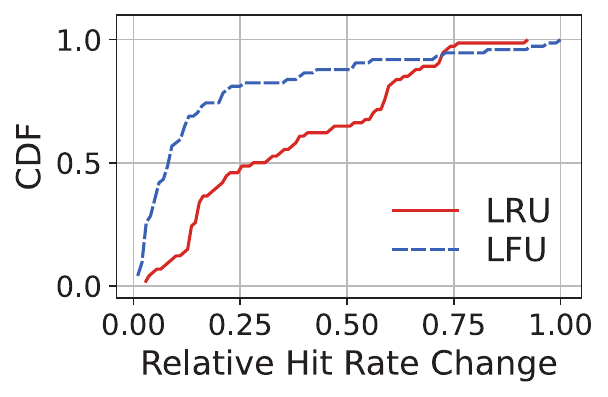}
        \label{fig:bg2}
    }
    \hspace{1mm}
    \subfloat[Hit rates under different number of concurrent clients.]{
    \includegraphics[width=0.465\columnwidth]{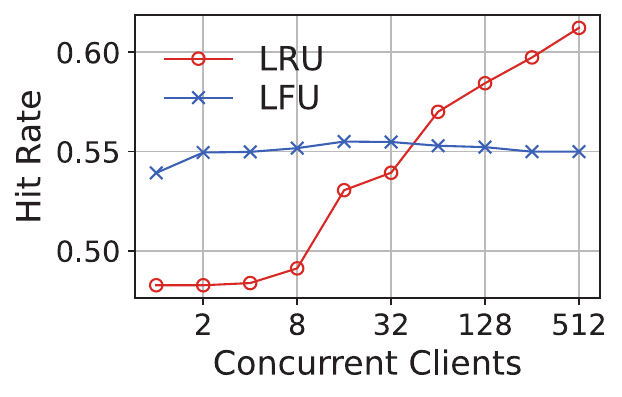}
        \label{fig:bg3}
    }
    \caption{The effect of concurrent clients on hit rates.}
    \label{fig:bg-client-num}
    \vspace{-0.1in}
\end{figure}

\textbf{\textit{1) Changing compute resources affects hit rates.}}
On caching systems on DM, applications execute multiple client threads on CPU cores in the compute pool to access cached data in the memory pool.
The access pattern on cached objects is the mixture of access patterns of all applications.
The change in compute resources, \ie the number of client threads of an application, alters the overall mixture of access patterns and affects the hit rate of individual caching algorithms in two ways.

First, the percentage of the data accesses of an application in the mixture changes with the number of client threads.
The overall access pattern on the cached objects thus changes since applications have dissimilar access patterns~\cite{atc17cidon}.
Figure~\ref{fig:bg1} shows the simulation result on a single machine with two applications under varying numbers of client threads.
One application executes an LRU-friendly workload and the other executes an LFU-friendly one from the FIU block trace~\cite{fast10koller}.
The hit rates of LRU and LFU are affected by the change of the compute resources in applications, where LFU exhibits a better hit rate when the LFU-friendly application has more compute resources and vice versa.

Second, the number of concurrent clients in an application changes the original access pattern of a workload due to concurrent executions.
We simulate on 74 real-world workloads from Twitter~\cite{osdi20yang} and FIU~\cite{fast10koller} with numbers of clients ranging from 1 to 512.
Figure~\ref{fig:bg2} shows the cumulative distribution function (CDF) of the relative hit rate change in these workloads. 
The relative hit rate change is calculated as $\frac{h_{max}-h_{min}}{h_{max}}$, where $h_{max}$ and $h_{min}$ are the highest and lowest hit rates of a workload under different numbers of clients.
As we increase the number of client threads, 80\% of workloads have 60\% hit rate change in LRU and 21\% in LFU.
Meanwhile, the best caching algorithms on 36\% of workload change with the varying number of concurrent clients.
Figure~\ref{fig:bg3} shows an example FIU trace where the hit rate of LFU performs better with a small number of concurrent clients but becomes inferior to LRU when the number of clients increases.

\textbf{\textit{2) Changing memory resources affects hit rates.}}
Changing memory resources leads to changing cache sizes of caching systems on DM.
For individual workloads, the best caching algorithm that maximizes the hit rate changes with cache sizes~\cite{fast21rodriguez}, \eg one workload can be LRU-friendly with a small cache size but becomes LFU-friendly under bigger cache sizes.
Our simulation finds that the best algorithm changes in 22 of the 74 real-world workloads when the cache size changes.
Figure~\ref{fig:bg4} shows an example FIU trace where LRU performs better with small caches and LFU performs better with larger cache sizes.

Consequently, it is necessary for caching systems on DM to dynamically select the best caching algorithm according to the changing resource settings.
However, achieving adaptivity is difficult on DM due to its decentralized and distributed nature, as we will introduce in \S~\ref{sec:adaptive-caching}.
\section{The \dmc Design}\label{sec:design}

\begin{figure}
    \centering
    \includegraphics[width=\columnwidth]{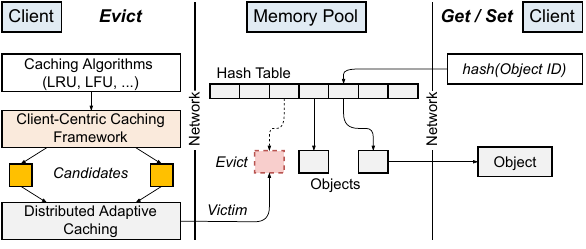}
    \caption{The overview of \dmc.}\label{fig:overview}
    \vspace{-0.1in}
\end{figure}

\subsection{Overview}
\noindent
Figure~\ref{fig:overview} shows the overall architecture of \dmc.
\dmc adopts a hash table to organize objects stored in the memory pool.
The hash table stores pointers to the addresses of the cached objects.
Following existing architectures of storage systems on DM~\cite{sigcomm21singhvi,fast23shen}, applications execute on CNs and each application owns a local \dmc client as a subprocess.
Each \dmc client has multiple threads on dedicated cores and applications communicate with \dmc clients with local shared memory to execute \textit{Get} and \textit{Set} operations.
Under this architecture, applications can freely scale compute resources by adding or removing the number of threads and CPU cores assigned to \dmc.
The adjustment on compute resources is independent against cached data because there is no need to increase or decrease the cache size in the memory pool when adding or reducing CPU cores.

\dmc clients execute \textit{Get} and \textit{Set} operations with one-sided RDMA verbs similar to RACE hashing~\cite{atc21zuo}, the state-of-the-art hashing index on DM.
For \textit{Get}s, a client searches the address of the cached object in the hash table and fetches the object from the address with two \textsf{RDMA\_READs}.
For \textit{Set}s, a client searches the slot of the cached object in the hash table with an \textsf{RDMA\_READ}, writes the new object to a free location with an \textsf{RDMA\_WRITE}, and atomically modifies the pointer in the slot with an \textsf{RDMA\_CAS}.

\dmc proposes a client-centric caching framework (\S~\ref{sec:sample-frmk}) and a distributed adaptive caching scheme (\S~\ref{sec:adaptive-caching}) to achieve cache eviction on DM.
The client-centric caching framework efficiently executes multiple caching algorithms on DM by selecting multiple eviction candidates of various caching algorithms.
The distributed adaptive caching scheme uses machine learning to learn the characteristics of the current data access pattern and evicts the candidate selected by the caching algorithm that performs the best.

\begin{table}[t]
    \vspace{0.1in}
    \newcommand{\tabincell}[2]{\begin{tabular}{@{}#1@{}}#2\end{tabular}}
    \footnotesize
    \centering
    \caption{The recorded access information.}
    \label{tab:acc-info}
    \begin{tabular}{c|l|c|c}
        \toprule[2pt]
        \textbf{Name}       & \textbf{Description}                                             & Global?    & Stateful?  \\
        \midrule[1pt]
        \textit{size}       & Object size                                                      & \Checkmark & \XSolidBrush    \\
        \textit{insert\_ts} & Insert timestamp                                                 & \Checkmark & \XSolidBrush    \\
        \textit{last\_ts}   & Last access timestamp                                            & \Checkmark & \XSolidBrush    \\
        \textit{freq}       & The number of accesses                                           & \Checkmark & \Checkmark \\
        \textit{latency}    & Access latency                                                   & \XSolidBrush    & \XSolidBrush    \\
        \textit{cost}       & \tabincell{l}{Cost to fetch the object\\from the storage server} & \XSolidBrush    & \XSolidBrush    \\
        \bottomrule[2pt]
    \end{tabular}
\end{table}

\subsection{Client-Centric Caching Framework}\label{sec:sample-frmk}
\noindent
The client-centric caching framework addresses the challenges of evaluating object hotness and selecting eviction candidates when executing caching algorithms on DM.

First, to assess the hotness of cached objects, \dmc records objects' access information and decides objects' hotness by defining and applying priority functions to the recorded access information.
Specifically, \dmc associates each object with a small metadata recording its global access information, \eg access timestamps, frequency, etc.
The metadata is updated collaboratively by clients with one-sided RDMA verbs after each \textit{Get} and \textit{Set}.
On the client side, \dmc offers two interfaces to integrate caching algorithms, \ie priority functions (\texttt{\small double priority(Metadata)}) and metadata update rules (\texttt{\small void update(Metadata)}).
A priority function maps the metadata of an object to a real value indicating its hotness.
Since the key difference between caching algorithms is their definition of object hotness, various caching algorithms can be integrated by defining different priority functions with the \texttt{\small priority} interface.
To support as many algorithms to be simply integrated with the \texttt{\small priority} interface as possible, we summarize the access information commonly used by existing caching algorithms~\cite{cusr03podlipnig} in Table~\ref{tab:acc-info} and record them in \dmc by default.

For advanced caching algorithms that require more access information, we allow algorithms to extend and define their own rules to update the metadata with the \texttt{\small update} interface.
Listing~\ref{lst:lru-k-eg} shows an example implementation of LRU-K~\cite{lru-k}.
LRU-K evicts objects with the smallest timestamp at its last K$^{th}$ access.
We split the $last\_ts$ field into K timestamps with lower precision and construct a ring buffer with the $freq$ counter.
If the object is accessed more than K times, then its priority is its last K$^{th}$ access timestamp, which is indexed by $(freq - K + 1)\text{ mod }K$.
Otherwise, we return the $insert\_ts$ of the object to achieve FIFO eviction in the access buffer~\cite{2q}.
We resort to storing the modified timestamp of LRU-K with cached objects if we need to simultaneously execute LRU-K with caching algorithms that rely on $last\_ts$, \eg LRU.

Second, to efficiently select eviction candidates of various caching algorithms on DM, \dmc adopts sampling with client-side priority evaluation.
The overhead of maintaining expensive caching data structures is then avoided.
Specifically, on each eviction, \dmc randomly samples $K$ objects in the cache and applies the defined priority functions to the access information of the sampled objects.
The eviction victim is approximated as the object with the lowest priority among $K$ sampled objects.

To efficiently execute the framework on DM, \dmc proposes a \textit{sample-friendly hash table} and a \textit{frequency-counter cache} to reduce the overhead of sampling objects and recording access information on DM.

\lstset{
    frame=single,
    xleftmargin=.1\columnwidth, xrightmargin=.1\columnwidth,
    belowskip=-1.5\baselineskip,
    aboveskip=-0.1\baselineskip
}
\begin{lstlisting}[float=tp, language=python, caption={The pseudocode of LRU-K.}, label={lst:lru-k-eg}, captionpos=t]
def update(Metadata m):
    m.freq += 1
    idx = m.freq % K
    m.last_ts[idx] = current_timestamp()

def priority(Metadata m):
    if m.freq < K:
        return m.insert_ts
    idx = (m.freq - K + 1) % K
    return m.last_ts[idx]
\end{lstlisting}

\subsubsection{Sample-friendly hash table}\label{sec:sample-hashtable}
\noindent
The sample-friendly hash table reduces the overhead of recording access information and sample objects on DM.
Specifically, sampling objects on DM suffers from high access latency because multiple \textsf{RDMA\_READs} are required to fetch the metadata of objects scattered in the memory pool.
Moreover, updating access information affects the overall throughput because these additional RDMA operations consume the bounded message rate of RNICs in the memory pool.

The sample-friendly hash table co-designs the sampling process with the hash index to address these two problems.
First, instead of storing all metadata together with objects, \dmc stores the most widely used metadata (\ie the default ones) together with the slots in the hash index but retains the metadata extensions required by advanced caching algorithms in objects.
With the co-designed hash table, sampling can be conducted with only one \textsf{RDMA\_READ} by directly fetching continuous slots with a random offset in the hash table.
Second, \dmc reduces the number of RDMA operations on updating object metadata by organizing access information according to their update frequency.
The well-organized access information enables multiple access information to be updated with a single \textsf{RDMA\_WRITE}.

\begin{figure}
    \centering
    \includegraphics[width=0.83\columnwidth]{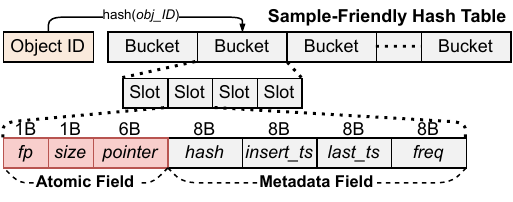}
    \caption{The sample-friendly hash table structure.}
    \label{fig:sf-ht}
\end{figure}

\textbf{Hash table structure.}
Figure~\ref{fig:sf-ht} shows the structure of the sample-friendly hash table.
The hash table has multiple buckets with multiple slots.
Each slot consists of two parts, \ie an atomic field and a metadata field.
The atomic field is similar to the slot of Race Hashing~\cite{atc21zuo}, which is 8-byte in length and modified atomically with \textsf{RDMA\_CAS}es when objects are inserted, updated, or deleted.
The atomic field contains a 6-byte \textit{pointer} referring to the address of the object, a 1-byte \textit{fp} (fingerprint) accelerating object searching, and a 1-byte \textit{size} recording the size of the stored object.
Similar to RACE hashing~\cite{atc21zuo}, we use a 1-byte size field and measure the sizes of objects in the granularity of 64B memory blocks. 
For large objects, \dmc stores the remaining data in a second memory block that links to the first one.
The metadata field records the access information required by most caching algorithms, as summarized in Table~\ref{tab:acc-info}. 
An additional \textit{hash} field is recorded for the distributed adaptive caching scheme, which will be discussed in \S~\ref{sec:adaptive-caching}.

\textbf{Access information organization.}
\dmc organizes the stored access information in two ways to reduce the number of RDMA operations on metadata updates.
First, \dmc reduces the number of access information that has to be included in the metadata by distinguishing local and global information.
Global information has to be maintained collaboratively by all clients and thus must be included in the metadata.
Local information can be decided locally by distributed clients and hence does not need to be included.
The \textit{latency} and \textit{cost} are local information because we assume that the latency and cost are approximately the same among clients and can be estimated based on the size of objects and the latency and cost of accessing other objects.
Second, global information is further classified into stateless and stateful information.
Stateless information is updated by overwriting its old value, while stateful information is updated based on its old value.
For instance, the \textit{insert\_ts} and \textit{last\_ts} are stateless because the old timestamps are no longer useful.
The \textit{freq} is stateful because it is always updated to increase by 1.
\dmc groups the stateless information together in the metadata so that they can be updated with a single \textsf{RDMA\_WRITE}.
The stateful information is updated with \textsf{RDMA\_FAA}s.

\subsubsection{Frequency-counter cache}\label{sec:fc-cache}
\noindent
A client-side frequency-counter (FC) cache is proposed to further reduce the overhead of updating metadata.
With the sample-friendly hash table, updating metadata still requires two RDMA operations, \ie an \textsf{RDMA\_WRITE} to update the stateless information and an \textsf{RDMA\_FAA} to update the stateful \textit{freq}. 
These RDMA operations consume the message rate of the RNIC and thus limit the overall throughput of \dmc.
Besides, executing \textsf{RDMA\_FAA} on DM is expensive due to the contention in the internal locks of RNICs~\cite{atc16kalia}.
The FC cache aims to reduce the number of \textsf{RDMA\_FAA} on metadata updates.

The FC cache stems from the idea of write-combining on modern processors~\cite{intel1998wc}.
In modern processors, several write instructions in a short time window are likely to target the same memory region, \eg a 64-byte cache line.
The write combining scheme adopts a buffer to absorb writes to the same region in a short time window and convert them into a single memory write operation to save memory bandwidths.

Similar to write-combining, \dmc employs an FC cache as the write-combining buffer.
The FC cache contains entries recording the object ID, the address of the slot in the hash table and the delta value of the counter. 
We track the insert time of each cache entry to ensure that the frequency counters in the memory pool do not lag too much.
Each time an object is accessed, its update to the frequency counter is buffered in the FC cache.
The update to the remote frequency counter is deferred until a cache entry is evicted.

There are two situations when an entry will be evicted from the FC cache.
First, if the space of the FC cache is full, an entry with the earliest insert timestamp will be evicted.
Second, if the buffered delta value of an object is greater than a threshold $t$, the entry will be evicted.
On entry eviction, the buffered counter value is added to the slot metadata with a single \textsf{RDMA\_FAA} according to the recorded slot address, reducing the number of \textsf{RDMA\_FAA} to up to ${1}/{t}$.

\begin{figure}
    \centering
    \includegraphics[width=0.94\columnwidth]{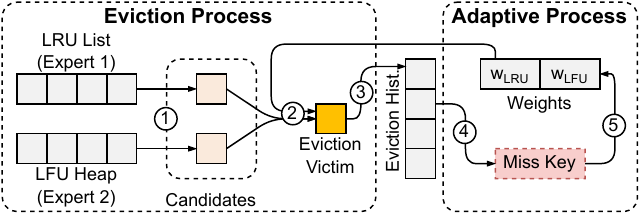}
    \caption{Adaptive caching on monolithic servers.}
    \label{fig:monolithic-adaptive}
\end{figure}

\subsection{Distributed Adaptive Caching}\label{sec:adaptive-caching}
\noindent
Adaptive caching on monolithic servers is proposed to adapt to changing data access patterns in real-world workloads.
\dmc proposes a distributed adaptive caching scheme to adapt to both changing workloads and dynamic resource settings on DM.
The key problem is how to achieve adaptive caching in a distributed and client-centric manner on DM.

Recent approaches on monolithic servers formulate adaptive cache as a multi-armed bandit (MAB) problem~\cite{fast21rodriguez,hotstorage18vietri,wdas02ari,fast03megiddon}.
As shown in Figure~\ref{fig:monolithic-adaptive}, caching servers simultaneously execute multiple caching algorithms, named experts in MAB~\cite{wdas02ari}.
Each expert is associated with a weight, reflecting its performance in the current workload.
The execution of the adaptive caching consists of an eviction and an adaptive process.
During the eviction process, each expert proposes an eviction candidate according to their own caching data structures (\textcircled{\small 1}).
Eviction victims are then decided opportunistically according to the weights of the experts (\textcircled{\small 2}), \ie candidates of experts with higher weights are more likely to be evicted.
The metadata of the evicted object, \ie the object ID and the experts choosing it as a candidate, are inserted into a fix-sized FIFO queue named eviction history (\textcircled{\small 3}).
During the adaptive process, existing approaches use \textit{regret minimization}~\cite{soda05flaxman,corr15foster,yusuf2020cache} to adjust expert weights.
Specifically, finding a missed object ID in the eviction history is a \textit{regret} because, intuitively, a more judicious eviction decision could have rectified the cache miss~\cite{hotstorage18vietri}.
Hence, when the missed object ID is found in the eviction history (\textcircled{\small 4}), the weights of experts deciding to evict the object are decreased (\textcircled{\small 5}).

Two challenges have to be addressed to achieve adaptive caching on DM.
First, maintaining the global FIFO eviction history is expensive due to the high overhead of accessing remote data structures on DM, as mentioned in \S~\ref{sec:challenges}. 
Second, managing expert weights on distributed clients is costly since clients need to be synchronized to get the updated weights.

The distributed adaptive caching scheme addresses these DM-specific challenges.
First, \dmc evaluates multiple priority functions with the client-centric caching framework to simultaneously execute multiple caching algorithms on DM.
Second, to avoid maintaining an additional FIFO queue on DM, \dmc embeds eviction history entries into the hash table with a lightweight eviction history (\S~\ref{sec:lw-history}).
Finally, to efficiently update and utilize expert weights on the client side, \dmc proposes a lazy weight update scheme to avoid the expensive synchronization among clients (\S~\ref{sec:async-weight-upd}).

\subsubsection{Lightweight eviction history}\label{sec:lw-history}
\noindent
The eviction history on monolithic servers needs to maintain an additional FIFO queue and an additional hash index to organize and index history entries~\cite{fast21rodriguez,hotstorage18vietri}.
The lightweight eviction history adopts two design choices to eliminate the overhead of maintaining these additional data structures on DM.
First, it uses an \textit{embedded history design} that reuses the slots of the sample-friendly hash table to store and index history entries.
No additional space needs to be allocated and no additional hash index needs to be constructed for history entries.
Second, the lightweight eviction history proposes a \textit{logical FIFO queue with a lazy eviction scheme} to efficiently achieve FIFO replacement on history entries.
No additional FIFO queue needs to be maintained to evict history entries.

\begin{figure}
    \centering
    \includegraphics[width=0.8\columnwidth]{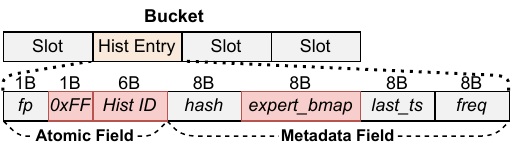}
    \caption{The structure of a lightweight history entry.}
    \label{fig:lw-hist-org}
\end{figure}

\textbf{Embedded history entries.}
Figure~\ref{fig:lw-hist-org} shows the structure of an embedded history entry of the lightweight history.
History entries are stored in the slots of the sample-friendly hash table with three differences.
First, the \textit{size} stores a special value (\textit{0xFF}) to tag the slot as a history entry.
We use \textit{0xFF} instead of 0 since we use 0 to indicate empty slots.
Second, the pointer field stores a 6-byte history ID instead of the address of the object.
Finally, the history entry uses the \textit{insert\_ts} of the slot to store a bitmap indicating which experts have decided to evict the object (\textit{expert\_bmap}).
Besides, each entry stores the hash value of the evicted object ID in the \textit{hash} field to check if a missed object is contained in the eviction history.
The hash value is written to the metadata when the object is inserted into the cache and will not be modified until its history entry is evicted from the FIFO eviction history.


\begin{figure}
    \centering
    \includegraphics[width=0.85\columnwidth]{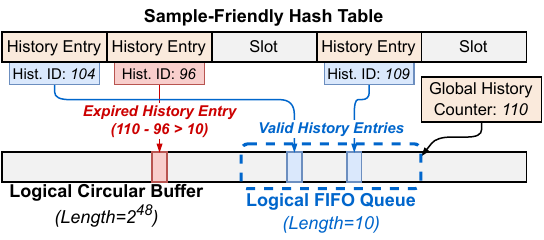}
    \caption{The logical FIFO queue structure.}
    \label{fig:logic-fifo}
\end{figure}

\textbf{The logical FIFO queue.}
The logical FIFO queue simulates FIFO eviction without actually maintaining a FIFO queue on DM.
It is constructed with a global history counter and the history IDs in history entries.
The global history counter is a 6-byte circular counter that generates history IDs for new history entries.
It is stored in an address in the memory pool known to all clients.
The history IDs of history entries are acquired by atomically reading the global history counter and increasing it by one (\ie atomic fetch-and-add).
As shown in Figure~\ref{fig:logic-fifo}, the global history counter and history IDs of history entries can be viewed as locations in a logical circular buffer with $2^{48}$ entries.
Combined with the size of the FIFO eviction history, the logical FIFO queue is then constructed, where the global history counter is the tail of the FIFO queue and the history IDs represent the location of history entries in the queue.

\begin{figure}
    \centering
    \includegraphics[width=0.9\columnwidth]{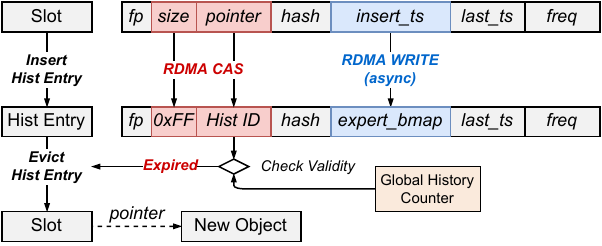}
    \caption{Inserting and evicting a history entry.}
    \label{fig:hist-insert-evict}
\end{figure}

Figure~\ref{fig:hist-insert-evict} shows the operations of the lightweight history:

\underline{\textit{History insertion.}}
A client inserts a history entry when it decides to evict a victim object from the cache.
The client first acquires a history ID by performing an \textsf{RDMA\_FAA} on the global history counter, which atomically returns the current value of the counter and increases it by one.
Then the client issues an \textsf{RDMA\_CAS} to atomically modify the \textit{size} and the \textit{pointer} in the slot of the victim object to be \textit{0xFF} and the acquired history ID, respectively. 
The expert bitmap is then asynchronously written to the \textit{insert\_ts} field of the slot metadata with an \textsf{RDMA\_WRITE}.

\underline{\textit{Lazy history eviction.}}
\dmc adopts a lazy eviction scheme to achieve FIFO eviction on history entries, \ie expired history entries are kept in the history for a while before their evictions.
To prevent clients from accessing expired history entries, \dmc proposes a client-side expiration checking mechanism.
Suppose the global history counter is $v_1$, the history ID is $v_2$, and the size of the FIFO history is $l$.
If $v_1 > v_2$, the history entry is invalid when $v_1 - v_2 > l$.
Otherwise, the history entry is invalid if $v_1 + 2^{48} - v_2 > l$, considering the wrap-up of the 48-bit global history counter.
The actual evictions happen when inserting new objects into the cache.
As shown in Figure~\ref{fig:hist-insert-evict}, when inserting new objects, the expired slots are considered empty slots and are overwritten to be ordinary slots, which transparently evicts the history entry.

\underline{\textit{Regret collection.}}
A regret is defined as a client finding an object to be missed in the cache but contained in the eviction history.
The embedded history entry makes collecting regrets the same process as searching objects in the cache.
When a client searches for an object, it calculates the hash value of the object ID, locates a bucket based on the hash value, and iteratively matches the slots in the bucket to see if the pointed object has the same object ID as the target.
During the process, clients also match the hash value of the encountered history entries in the bucket.
Regrets can then be collected if the object has not been found but a history entry has a matching hash value.

\subsubsection{Lazy expert weight update}\label{sec:async-weight-upd}
\noindent
\dmc formulates the problem of cache replacement as MAB and uses regret minimization to dynamically adjust the weights of experts.
When a regret is found, \ie a missed object hits in the eviction history, the weights of the experts that evicted the object should be penalized.
Suppose expert $E_i$ made a bad eviction decision and the decision is the $t$-th entry in the eviction history.
The weight of the expert is then updated to be $w_{E_i} = w_{E_i} \cdot e^{\lambda * d^t}$, where $\lambda$ is the learning rate and $d^t$ is the penalty.
The penalty $e^{\lambda * d^t}$ is related to the position of the entry in the FIFO history because an older regret should be penalized less, where $d$ is a fixed discount rate\footnote{Similar to~\cite{hotstorage18vietri}, the discount rate is $0.005^{1/N}$, where $N$ is the cache size.}.
The challenge of updating weights on DM is that regrets are no longer collected and expert weights are no longer used in a centralized manner by monolithic caching servers.
Updating and using expert weights from distributed clients incurs nonnegligible overhead due to the high synchronization overhead on DM~\cite{atc20tsai}.

The idea of the lazy weight update scheme is to let clients batch the regrets locally and offload the weight update lazily to the controllers of MNs.
In this way, the frequency of updating weights is reduced and the overhead of synchronization is avoided.
Meanwhile, the weak controller of memory nodes will not become a bottleneck due to the infrequent update.

\begin{figure}
    \centering
    \includegraphics[width=0.9\columnwidth]{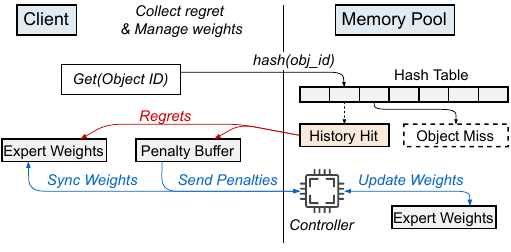}
    \caption{The lazy weight update scheme.}\label{fig:async-weigh-upd}
\end{figure}

Figure~\ref{fig:async-weigh-upd} shows the process of the lazy expert weight update scheme.
Each client maintains expert weights locally to make eviction decisions.
When a client discovers a regret, it applies the penalty to the local expert weights according to the history bitmap in the history entry.
The penalties are recorded in a penalty buffer.
When the number of buffered penalties exceeds a threshold, the client sends all the penalties to the controller of the memory node holding the expert weights with an RDMA-based RPC request.
On receiving clients' penalties, the controller of the MN first applies the penalties to the global expert weights and then replies the updated global weights to clients.

To reduce the bandwidth consumption of transferring the penalties over the network, \dmc compresses the penalties using the attribute of exponential functions.
Specifically, the sum of the penalties is stored in the penalty buffer and transferred to the MN instead of a list of individual penalties.

With the lazy weight update scheme, clients' eviction decisions are made on local weights, which are not always synchronized with global weights.
However, such asynchrony does not affect the adaptivity of \dmc, as shown in our experiments.

\subsection{Discussions}
\noindent\textbf{\textit{Metadata extensions.}}
As mentioned in \S~\ref{sec:sample-hashtable}, \dmc stores extended metadata together with cached objects for advanced caching algorithms.
In this situation, the extended metadata is stored as a metadata header ahead of each object.
The \texttt{update} and \texttt{priority} functions take all metadata, \ie the default ones in the hash table and the extended ones in the metadata header, as input and call user-defined metadata update and priority calculation rules to deal with the extended metadata.
After executing \textit{Get} and \textit{Set} operations, an additional \textsf{RDMA\_WRITE} is required to update the metadata stored with objects asynchronously.
Finally, on cache eviction, additional \textsf{RDMA\_READ}s are required to fetch the metadata header to calculate eviction priorities.

\noindent
\textbf{\textit{Metadata overhead.}}
In \dmc, metadata consists of history entries, the index slots for cached objects and global expert weights.
First, each history entry contains 40 bytes, as shown in Figure~\ref{fig:lw-hist-org}.
The total number of history entries is set as the maximum number of cached objects according to existing approaches~\cite{hotstorage18vietri,fast21rodriguez}.
Second, for each cached object, the index slot uses 40 bytes, \ie 8 bytes for the atomic field and 32 bytes for access information, as shown in Figure~\ref{fig:sf-ht}.
Finally, for each expert, a 4-byte float variable is required as its global expert weight.
Summing up all of these, the metadata overhead of \dmc is $80\cdot C + 4 \cdot N$ bytes, where $C$ is the maximum number of cached objects and $N$ is the number of experts in the distributed adaptive caching scheme.

\noindent
\textbf{\textit{Security and fairness issues.}}
Since \dmc clients and applications cooperate closely on the same CNs, it is possible that some malicious users can manipulate \dmc clients to make them disproportionately advantaged against other users' applications.
We can enforce security techniques, \eg control flow integrity (CFI)~\cite{cfi}, on standalone \dmc clients to prevent \dmc clients from being manipulated.
We can also integrate the expected delaying technique~\cite{nsdi16pu} in \dmc clients to ensure that applications fairly share the cache.
\section{Evaluation}\label{sec:eval}
\noindent
The evaluation of \dmc answers the following questions:
\begin{itemize}
    \item \textbf{Q1:} How elastic is \dmc compared with caching systems on monolithic servers?
    \item \textbf{Q2:} How efficient is \dmc in executing caching algorithms on DM?
    \item \textbf{Q3:} How adaptive is \dmc to real-world workloads and the changing resources on DM?
    \item \textbf{Q4:} How flexible is \dmc in integrating various caching algorithms on DM?
    \item \textbf{Q5:} How does each design point contribute to \dmc?
\end{itemize}

\subsection{Experimental Setup}
\noindent\textbf{\textit{Testbed.}}
We evaluate \dmc with 9 physical machines (8 CNs and 1 MN) on the Clemson cluster of CloudLab~\cite{atc19duplyakin}.
Each machine has two 36-core Intel Xeon CPUs, 256 GB DRAM, and a 100Gbps Mellanox ConnectX-6 NIC.
All machines are connected to a 100Gbps Ethernet switch.
In all our experiments, we use a single physical machine and use one CPU core to simulate the memory pool of DM with weak compute power~\cite{atc20tsai,fast23shen}.
\dmc is compatible with memory pools with multiple MNs as long as the memory pool offers the required interfaces presented in \S 2.2.
Besides, we use up to 32 cores on CNs, with each executing a client thread because they are on the same NUMA node with the RNIC.

\begin{table}[t]
    \vspace{0.15in}
    \footnotesize
    \centering
    \caption{\small Real-world workloads used in the evaluation.}
    \label{tab:wl-info}
    \begin{tabular}{l|l|r}
        \toprule[2pt]
        \textbf{Workload} & \textbf{Workload Type} & \textbf{\# Requests}            \\
        \midrule[1pt]
        \textit{IBM}               & Object Store              & 10 - 40 million  \\
        \textit{CloudPhysics}      & Block IO                  & 50 million       \\
        \textit{Twitter-Transient} & Transient key-value cache & 10 million       \\
        \textit{Twitter-Storage}   & Storage key-value cache   & 10 million       \\
        \textit{Twitter-Compute}   & Compute key-value cache   & 10 million       \\
        \textit{webmail}           & Block IO                  & 7.8 million      \\
        \bottomrule[2pt]
    \end{tabular}
\end{table}

\noindent\textbf{\textit{Workloads.}}
We evaluate \dmc with both YCSB synthetic workloads~\cite{socc10cooper} and real-world key-value traces~\cite{nsdi20song,osdi20yang,fast10koller}. 
For YCSB synthetic workloads, we use 4 core workloads: A (50\% GET, 50\% UPDATE), B (95\% GET, 5\% UPDATE), C (100\% GET), and D (95\% GET, 5\% INSERT). 
For all four workloads, we pre-generate 10 million keys with 256-byte key-value pairs, load these generated keys by sharding them to all clients, and execute the corresponding workloads.
The requests are generated with Zipfan distribution with $\theta=0.99$.
For real-world key-value traces, we use workloads from IBM~\cite{hotstorage20eytan}, CloudPhysics~\cite{fast15waldspurger}, Twitter~\cite{osdi20yang}, and FIU~\cite{fast10koller}, as shown in Table~\ref{tab:wl-info}.
The \textit{IBM} trace is collected from IBM Cloud Object Storage~\cite{hotstorage20eytan}.
We ignore traces with less than 10 million requests since they have too few unique objects and use all 23 traces in our experiments.
The \textit{CloudPhysics} dataset includes block I/O traces on VMs with different CPU/DRAM configurations~\cite{fast15waldspurger}.
We use the first 10 traces with more than 50 million requests to evaluate \dmc.
For the Twitter traces, we randomly select three traces, \ie \textit{Twitter-Compute}, \textit{Twitter-Storage}, and \textit{Twitter-Transient}, from a compute cluster, a storage cluster, and a transient caching cluster, respectively.
The \textit{webmail} trace is a 14-day storage I/O trace collected from web-based email servers.
We use \textit{webmail} as a representative FIU trace similar to existing approaches~\cite{fast21rodriguez}.
In our experiments, we randomly select traces to accelerate our evaluation to show the performance of \dmc in different use cases, \ie block IO, KV cache on different clusters, and object store. 
We truncate traces to allow concurrent trace loading from 32 independent clients on a single CN.

\noindent\textbf{\textit{Implementations.}}
We implement \dmc with 20k LOCs.
We use LRU and LFU, the two most widely used caching algorithms, as two experts in the distributed adaptive caching scheme.
These two caching algorithms are chosen as adaptive experts since existing adaptive caching schemes have found that using a recency-based and a frequency-based caching algorithm can adapt to most workloads~\cite{fast21rodriguez,hotstorage18vietri}.
For memory management, we use a two-level memory management scheme~\cite{fast23shen} so that clients can dynamically allocate memory spaces in the MN.
We pre-register all memory on the MN to its RNIC to eliminate the overhead of memory registration on the critical path of memory allocation.

\noindent\textbf{\textit{Parameters.}}
The parameters of \dmc include the number of samples, the size of lightweight eviction history, the threshold and size of the FC Cache, and the learning rate and the number of batched weight updates of distributed adaptive caching.
Specifically, the number of samples affects the precision of approximating caching algorithms with sampling.
We sample 5 objects on cache eviction according to the default value of Redis~\cite{redis}.
The size of the lightweight eviction history exhibits a tradeoff between the speed of adaptation and the metadata overhead.
Setting the history size larger makes adaptation faster since more penalties can be collected during execution.
In return, a larger history size requires more space to store history entries.
We set the history size as the cache size (calculated in the number of objects) according to LeCaR~\cite{hotstorage18vietri}.
The threshold of FC Cache can affect the precision of LFU. 
We set the FC cache threshold to 10 and set the FC cache size to 10MB according to our grid search.
The superior hit rates in our experiments show that using 10 as the FC threshold does not affect hit rates much.
Finally, we configure the learning rate of \dmc to be 0.1 and update global weights every 100 local weight updates according to our grid search.

\noindent\textbf{\textit{Baselines.}}
We compare \dmc with Redis~\cite{redis}, CliqueMap~\cite{sigcomm21singhvi}, and Shard-LRU.
First, we use Redis, one of the most widely adopted in-memory caching systems that support dynamic resource scaling~\cite{redis,elasticache}, to show the elasticity of \dmc.
Second, we use CliqueMap, the state-of-the-art RDMA-based KV cache from Google, to show the efficiency and adaptivity of \dmc.
CliqueMap initiates \textsf{RDMA\_READ}s on the client side to directly \textit{Get} cached objects, and relies on server-side CPUs to execute \textit{Set} operations.
Since \textit{Get}s involves only one-sided \textsf{RDMA\_READ}s, no access information can be recorded.
Clients of CliqueMap record access information locally and send the information to servers periodically to enable servers to execute caching algorithms.
We implemented an LRU (CM-LRU) and LFU (CM-LFU) version of CliqueMap according to its paper due to no open-source implementations.
We disable the replication and fault-tolerance of CliqueMap to focus on comparing the execution of caching algorithms.
Finally, we use Shard-LRU, a straightforward implementation of a caching system on DM, to show the effectiveness of the client-centric caching framework of \dmc.
Clients of Shard-LRU maintain lock-protected LRU lists in the memory pool with one-sided RDMA verbs.
We shard objects into 32 LRU lists according to their hash values and force clients to sleep 5 us on lock failures to mitigate lock and network contention.
By default, we use one CPU core on MNs to simulate the poor compute power in the memory pool.
Each CPU core on CNs exclusively runs a client thread.

\subsection{Q1: Elasticity}
\noindent
To show the elasticity of \dmc, we run the same experiment as in \S~\ref{sec:background} and force \dmc to use the same amount of CPU or memory resources as Redis on the YCSB-C workload.

\begin{figure}
    \centering
    \includegraphics[width=\columnwidth]{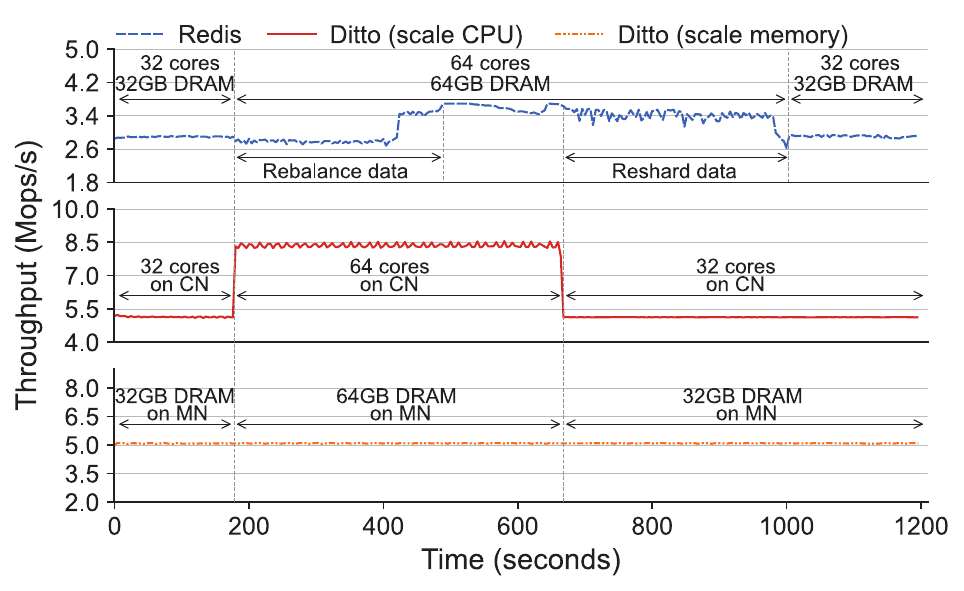}
    \caption{The throughput of \dmc when dynamically adjusting compute and memory resources.}
    \label{fig:ditto-elasticity-tpt}
    \vspace{-0.1in}
\end{figure}

\begin{figure*}
    \centering
    \subfloat[YCSB A]{
        \includegraphics[width=0.5\columnwidth]{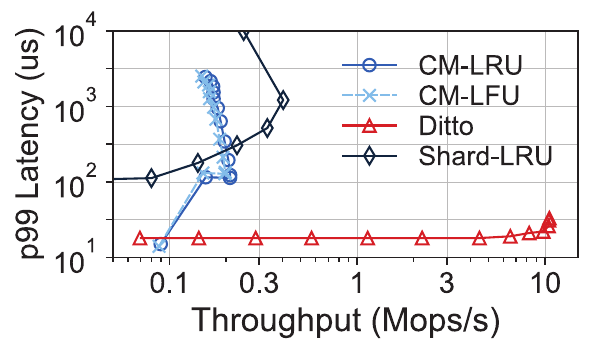}
        \label{fig:ycsb-a}
    }
    \subfloat[YCSB B]{
        \includegraphics[width=0.5\columnwidth]{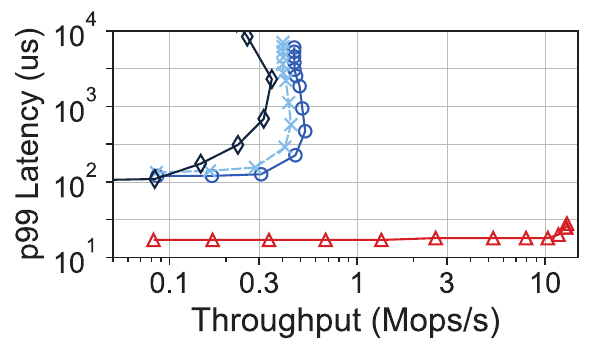}
        \label{fig:ycsb-b}
    }
    \subfloat[YCSB C]{
        \includegraphics[width=0.5\columnwidth]{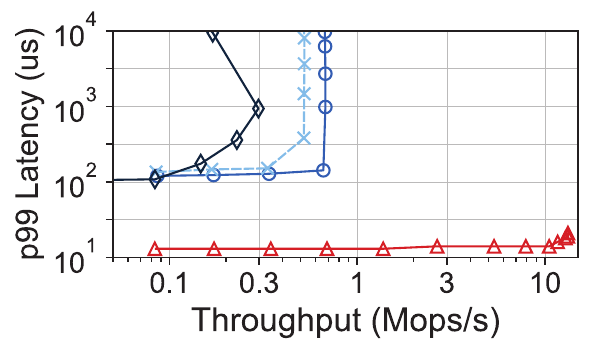}
        \label{fig:ycsb-c}
    }
    \subfloat[YCSB D]{
        \includegraphics[width=0.5\columnwidth]{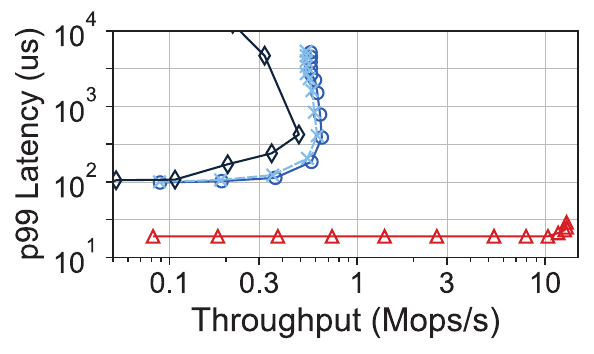}
        \label{fig:ycsb-d}
    }
    \caption{The throughput and tail latency of caching systems on DM.}
    \label{fig:ycsb-tpt-lat}
\end{figure*}

\begin{figure}
    \vspace{-0.2in}
    \centering
    \subfloat[YCSB A (write-intensive)]{
        \includegraphics[width=0.475\columnwidth]{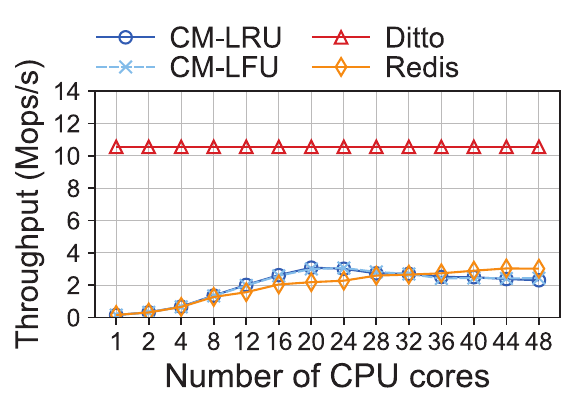}
        \label{fig:cm-core-ycsba}
    }
    \subfloat[YCSB C (read-only)]{
        \includegraphics[width=0.475\columnwidth]{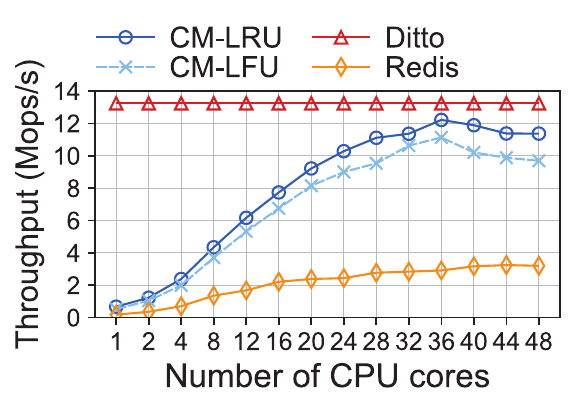}
        \label{fig:cm-core-ycsbc}
    }
    \caption{The throughput of CliqueMap, Redis, and \dmc with more CPU cores on MN.}
    \label{fig:cm-core}
    \vspace{-0.2in}
\end{figure}
Compared with Redis, the elasticity of \dmc is improved in both resource utilization and speed of resource adjustments.
First, due to the decoupled CPU and memory on DM, \dmc can adjust CPU cores and memory spaces separately in a fine-grained manner.
Resources can be allocated precisely according to the dynamic demands of applications.
Second, \dmc does not require data migration when adjusting resources, making the performance gain and resource reclamation more agile than Redis.
The throughput of \dmc improves immediately from 5 Mops to 8.5 Mops with 32 more CPU cores added and resumes immediately back to 5 Mops as we shrink the number of CPU cores back to 32.
The throughput doesn't scale linearly as we add CPU cores due to the extra overhead of coroutine scheduling on CNs.
The median latency stabilizes at 12 us and the 99th percentile latency fluctuates slightly around 14 to 21 us.
As for adjusting memory spaces, the throughput stabilizes on 5 Mops and the tail latency stays on 14 us.
Besides, the throughput of \dmc is more than 2 times higher than that of Redis during the entire experiment. 
This is because \dmc allows CPU cores to equally access all data, avoiding a single core becoming the performance bottleneck.
However, Redis shards data to VMs, which makes the CPU core of some VMs bottleneck the throughput of the entire caching cluster on the skewed YCSB workloads. 

Besides, \dmc does not require more client-side computation than Redis.
In the experiment, clients of \dmc consume 32 CPU cores on the CN.
In contrast, clients of Redis consume on average 36.3 CPU cores out of 128 assigned cores on two CNs.
This is because the Redis client library spends CPU cycles to encapsulate and decapsulate data according to the Redis communication protocol and network protocols.
Moreover, \dmc saves compute power regarding the overall CPU utilization since Redis servers consume an additional 32 cores on the MN.

\subsection{Q2: Efficiency}
To show that \dmc can efficiently execute caching algorithms on DM, we evaluate the throughput and tail latency of Shard-LRU, CliqueMap, and \dmc in the case of no cache misses on YCSB benchmarks.
We vary the number of clients from 1 to 256, with each CN holding up to 32 clients.

As shown in Figure~\ref{fig:ycsb-tpt-lat}, Shard-LRU is bottlenecked by its remote lock contention even if the sharded LRU list and the 5 us back-off scheme mitigate the lock and network contention.
The throughput of CliqueMap is limited by the weak compute power on MNs.
For write-intensive workloads (YCSB A), the CPU of the MN is overwhelmed by frequent \textit{Set}s.
For read-intensive workloads (YCSB B, C, and D), the CPU of the MN is busy with merging the object access information received from clients. 
The overall performance is affected by the periodic synchronization of access information and the amplified network bandwidth when sending the access information from clients to the MN.

For all workloads, \dmc is bottlenecked by the message rate of the RNIC on the MN.
It achieves 10.5, 13.1, 13.2, and 13.0 Mops respectively on YCSB A, B, C, and D workloads, which is up to $9\times$ higher than Shard-LRU and CliqueMap.
Compared with Shard-LRU, \dmc records the access information and selects eviction victims in a lock-free manner, eliminating the expensive lock overhead on DM.
Compared with CliqueMap, \dmc accesses data and maintains access information with one-sided RDMA verbs, preventing the weak compute power on the MN from becoming the throughput bottleneck on both write-intensive and read-intensive workloads.
However, \dmc performs worse than CliqueMap under the write-intensive YCSB-A workload with a single client, \ie the first point in Figure~\ref{fig:ycsb-a}.
This is because the \textit{Set}s of CliqueMap use only a single RTT, while \dmc needs three RTTs to search the remote hash table, read the object, and modify the pointer in the hash table.

\begin{figure*}
    \centering
    \subfloat[Webmail] {
        \includegraphics[width=0.4\columnwidth]{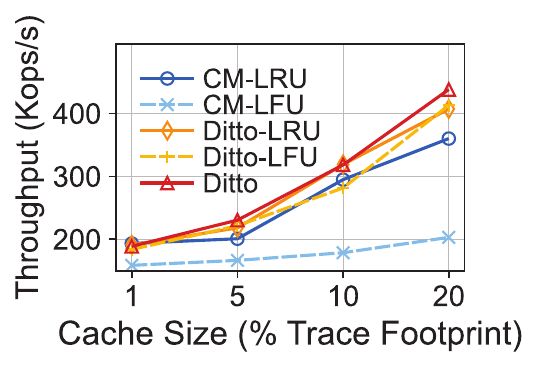}
        \label{fig:webmail-tpt}
    }
    \subfloat[Twitter-Transient]{
        \includegraphics[width=0.4\columnwidth]{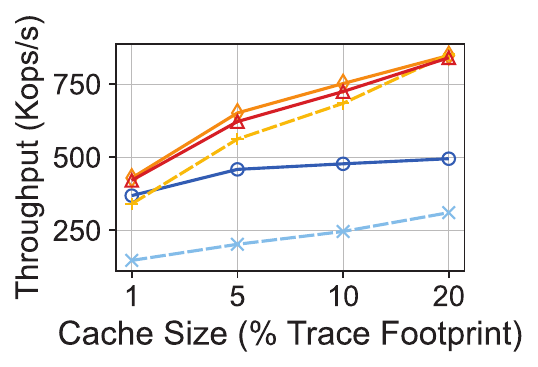}
        \label{fig:twitter-transient-tpt}
    }
    \subfloat[Twitter-Storage]{
        \includegraphics[width=0.4\columnwidth]{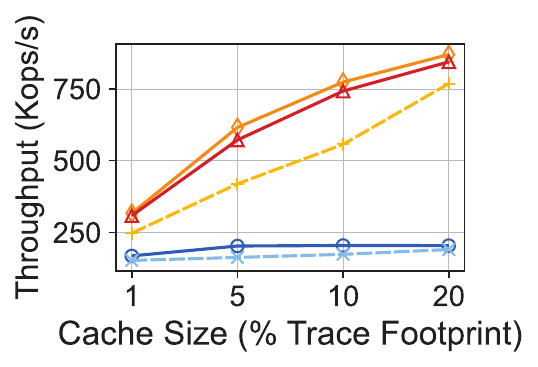}
        \label{fig:twitter-storage-tpt}
    }
    \subfloat[Twitter-Compute]{
        \includegraphics[width=0.4\columnwidth]{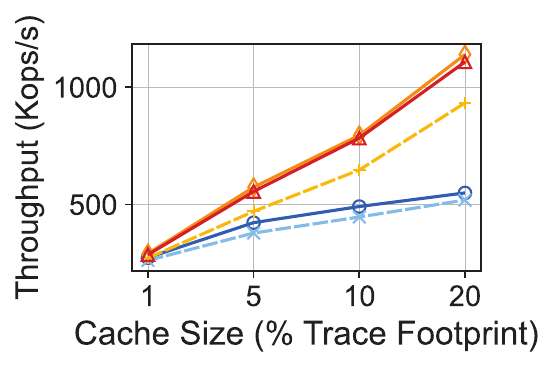}
        \label{fig:twitter-compute-tpt}
    }
    \subfloat[IBM]{
        \includegraphics[width=0.4\columnwidth]{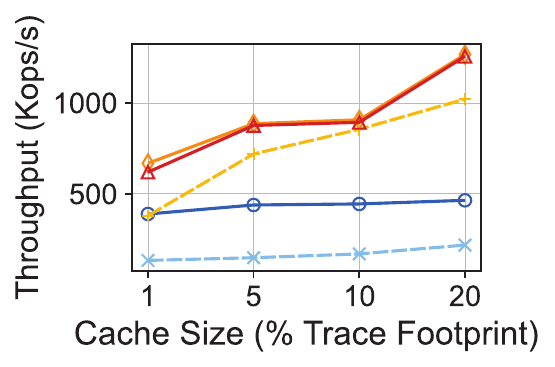}
        \label{fig:ibm-tpt}
    }
    \caption{Penalized throughputs under different real-world workloads.}
    \label{fig:real-tpt}
\end{figure*}

\begin{figure*}
    \centering
    \subfloat[Webmail] {
        \includegraphics[width=0.4\columnwidth]{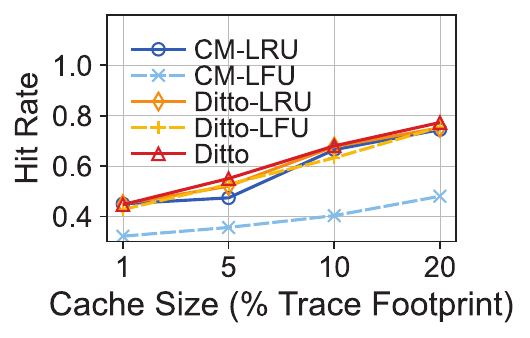}
        \label{fig:webmail-hr}
    }
    \subfloat[Twitter-Transient]{
        \includegraphics[width=0.4\columnwidth]{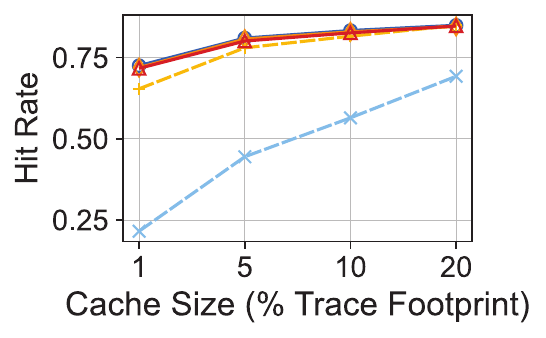}
        \label{fig:twitter-transient-hr}
    }
    \subfloat[Twitter-Storage]{
        \includegraphics[width=0.4\columnwidth]{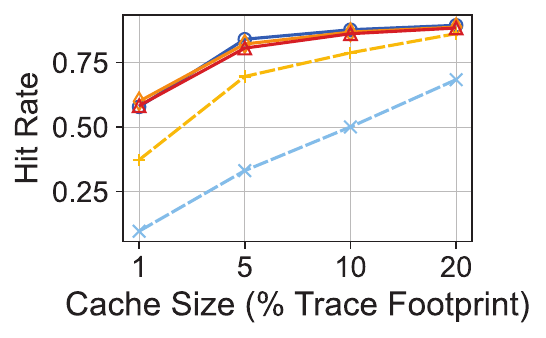}
        \label{fig:twitter-storage-hr}
    }
    \subfloat[Twitter-Compute]{
        \includegraphics[width=0.4\columnwidth]{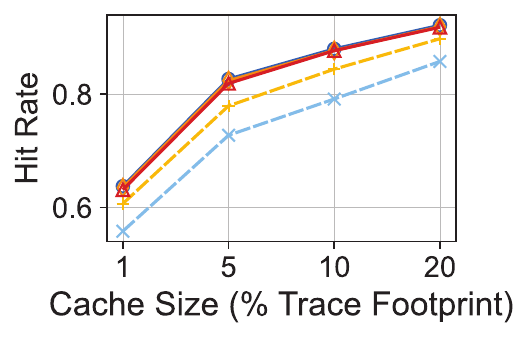}
        \label{fig:twitter-compute-hr}
    }
    \subfloat[IBM]{
        \includegraphics[width=0.4\columnwidth]{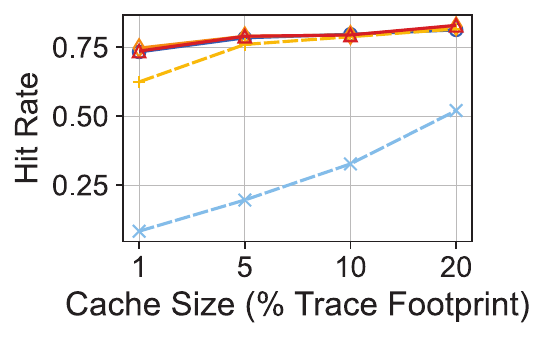}
        \label{fig:ibm-hr}
    }
    \caption{Hit rates under different real-world workloads.}
    \label{fig:real-hr}
\end{figure*}

Figure~\ref{fig:cm-core} shows the performance of CliqueMap, Redis and \dmc under YCSB-A and YCSB-C workloads with increasing numbers of MN-side CPU cores under 256 clients.
We shard the LRU list (and the LFU heap) of CliqueMap into 128 shards to avoid server-side lock contention.
The throughput of \dmc stays the same since \dmc does not rely on compute power on MNs.
With the same compute resource in the compute pool, CliqueMap consumes more than 20 additional cores to get comparable performance with \dmc on YCSB-C.
\dmc achieves $3.3\times$ higher throughput than CliqueMap on the write-intensive YCSB-A workload since CliqueMap relies only on the server-side compute power to execute \textit{Set} operations and maintain caching data structures.
The throughputs of Redis on both workloads are bottlenecked by the CPU core of the hottest data shard due to the skewed YCSB workloads.
Redis performs slightly better than CliqueMap on YCSB-A workload with more CPU cores since its sample-based eviction eliminates the overhead of maintaining caching data structures locally.

\subsection{Q3: Adaptivity}
\subsubsection{Adapt to real-world workloads}
\noindent
To show the \pagebreak adaptivity of \dmc on real-world workloads with different affinities of caching algorithms, we evaluate the throughput and the hit rate of real-world workloads with different cache sizes.
For all traces, we use 256-byte object sizes and set cache sizes relative to the size of each workload's footprint, \ie all unique data items accessed, similar to~\cite{fast21rodriguez}.
For each workload, we use 64 clients to first execute 10 seconds to warm up the cache and then let all clients iteratively run the workload for 20 seconds to calculate the hit rate and the throughput.
We use a penalized throughput to simulate real-world situations where caching systems cooperate with a distributed storage system.
For each \textit{Get} miss, we force clients to sleep for 500 us before inserting the missed object into the cache with \textit{Set}.
The penalty simulates the overhead of fetching data from distributed storage services and 500 us is selected according to the latency of the state-of-the-art distributed storage systems~\cite{osdi06weil,fast22lv,fast21pan}.

We compare \dmc with four baseline approaches.
We use CM-LRU and CM-LFU to show the performance of precise LRU and LFU implementation with CliqueMap on DM.
We introduce \dmc-LRU and \dmc-LFU to show the performance of \dmc with only a single caching algorithm.

Since \dmc is an adaptive caching framework that can execute various caching algorithms and dynamically adapt to the best one based on workloads and resource settings, the performance of \dmc largely depends on the candidate caching algorithms configured by users.
We configure \dmc to execute LRU and LFU as examples to show its adaptivity.
Under workloads that are friendly to either LRU or LFU, the performance of \dmc should be bounded by \dmc-LRU and \dmc-LFU and approach to the better one since it adaptively selects the better one among the two algorithms.

Figures~\ref{fig:real-tpt} and~\ref{fig:real-hr} show the penalized throughput and the hit rates under five real-world key-value traces.
In all five workloads, the hit rate and penalized throughput of \dmc can effectively approach the better one of \dmc-LRU and \dmc-LFU.
Meanwhile, \dmc outperforms CliqueMap in all workloads due to higher hit rates and the higher throughput upper-bound.
Particularly, the throughput of CliqueMap is bounded by the compute power on the MN under the Twitter workloads, where the hit rates are high.
One exception is the throughput of CM-LRU in Figure~\ref{fig:webmail-tpt}, which has comparable throughput with \dmc. 
This is because all approaches are bounded by the hit rate on the \textit{webmail} workload and CM-LRU has a slightly lower hit rate compared with \dmc.
For most of the workloads, the throughput of \dmc is lower than that of \dmc-LRU when their hit rates are the same due to the additional overhead of adaptive caching, \ie accessing and increasing the global history counter.
However, the overhead is less than 5\%, which is acceptable compared with the up to 63\% performance gain of using an inferior caching algorithm, since users do not know in advance which caching algorithm performs better.

\begin{figure}
    \begin{minipage}[t]{0.48\columnwidth}
        \centering
        \includegraphics[width=0.94\columnwidth]{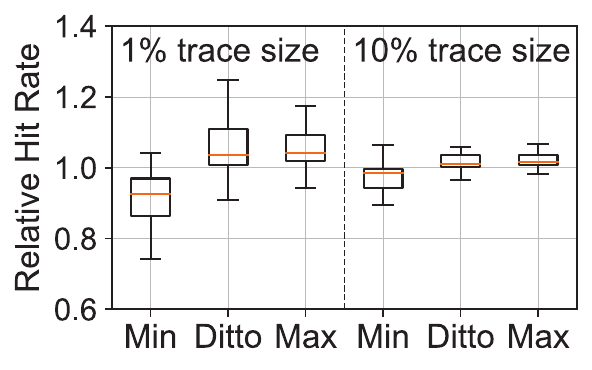}
        \caption{The relative hit rate of \dmc, \dmc-LRU, and \dmc-LFU on 33 workloads.}
        \label{fig:ditto-box}
    \end{minipage}%
    \hspace{2mm}
    \begin{minipage}[t]{0.48\columnwidth}
        \centering
        \includegraphics[width=1.1\columnwidth]{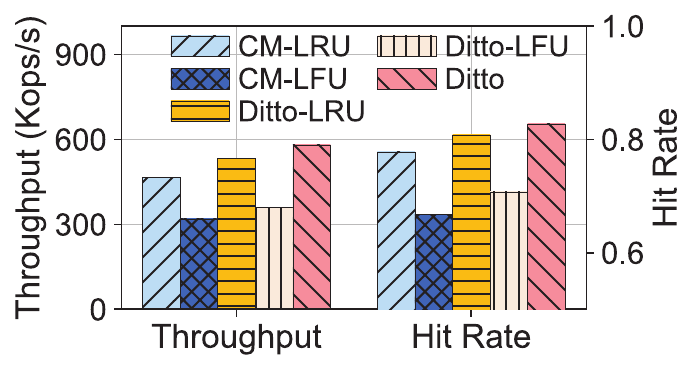}
        \caption{The penalized throughput and hit rate under a changing workload.}
        \label{fig:changing-tpt-hr}
    \end{minipage}
\end{figure}

\begin{figure}
    \begin{minipage}[t]{0.45\columnwidth}
        \centering
        \includegraphics[width=\columnwidth]{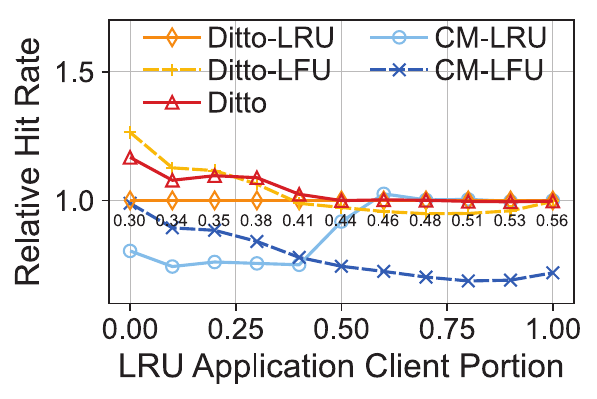}
        \caption{The relative hit rates under different proportions of clients assigned to LRU and LFU applications.}
        \label{fig:mix-hr}
    \end{minipage}
    \hspace{2mm}
    \begin{minipage}[t]{0.45\columnwidth}
        \centering
        \includegraphics[width=\columnwidth]{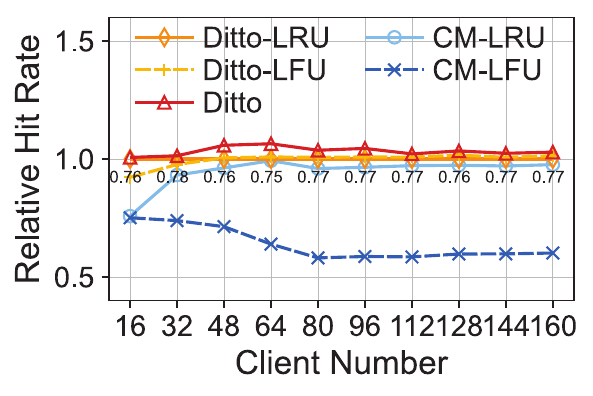}
        \caption{The relative hit rates of \dmc and CliqueMap when dynamically adding the number of concurrent clients.}
        \label{fig:para-hr}
    \end{minipage}
\end{figure}

Figure~\ref{fig:ditto-box} shows the box plot of relative hit rates of \dmc, max(\dmc-LRU, \dmc-LFU), and min(\dmc-LRU, \dmc-LFU) normalized over random eviction on 33 \textit{IBM} and \textit{CloudPhysics} workloads.
The hit rate of \dmc significantly exceeds min(\dmc-LRU, \dmc-LRU) and approaches the box of max(\dmc-LRU, \dmc-LFU), showing the adaptivity of \dmc.

Under changing workloads that iteratively switch between LRU- and LFU-friendly, \dmc should outperform both \dmc-LRU and \dmc-LFU.
We show the performance of the four approaches on a synthetic changing workload used in~\cite{hotstorage18vietri}.
The workload is synthesized to have four phases that periodically switch back and forth from being favorable to LRU to being favorable to LFU.
As shown in Figure~\ref{fig:changing-tpt-hr}, \dmc outperforms all baselines on both penalized throughput and hit rate because only \dmc can adapt to workload changes.

\subsubsection{Adapt to dynamic resource adjustments}
\noindent
To show the adaptivity of \dmc on DM, we evaluate its hit rates with dynamically changing compute and memory resources on the same workload as Figures~\ref{fig:bg1}, \ref{fig:bg4}, and~\ref{fig:bg3}, \ie \textit{webmail}.

\noindent\textbf{Adapt to changing compute resources.}
Figure~\ref{fig:mix-hr} shows the relative hit rates normalized to \dmc-LRU under different proportions of clients allocated to two applications with LRU and LFU access patterns.
The hit rate of \dmc-LFU is higher when the LRU portion is less than $0.4$, while \dmc-LRU performs better when the LRU portion grows higher.
The hit rate of \dmc is higher than that of \dmc-LRU with a low LRU portion and becomes close to \dmc-LRU with a high LRU portion, indicating the adaptivity of \dmc.
Besides, \dmc can adapt to the change of access pattern when multiple clients concurrently execute the same workload.
Figure~\ref{fig:para-hr} shows the relative hit rates of \dmc and CliqueMap normalized to \dmc-LRU under dynamically increasing numbers of concurrent clients\footnote{The absolute hit rates in Figures~\ref{fig:ditto-box}, \ref{fig:mix-hr}, and \ref{fig:para-hr} can be found in our open-source repository.}.
The hit rate of \dmc stays above the hit rates of both \dmc-LRU and \dmc-LFU because there are access pattern changes in the real-world \textit{webmail} workload, and only \dmc can adapt to these changes.

\begin{figure}
    \begin{minipage}[t]{0.47\columnwidth}
        \centering
        \includegraphics[width=\columnwidth]{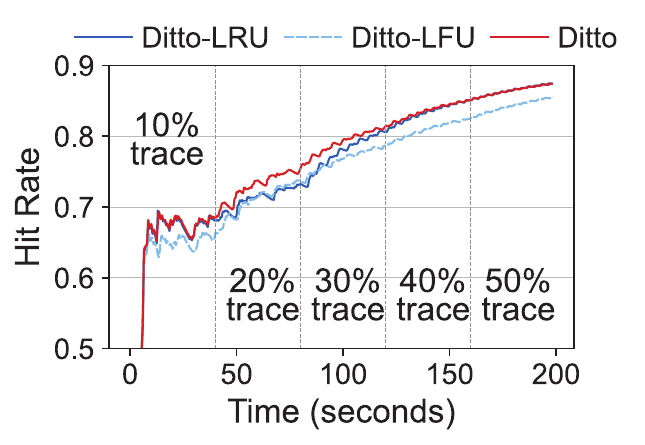}
        \caption{The hit rate under dynamic cache sizes.}
        \label{fig:mem-hr}
    \end{minipage}
    \hspace{2mm}
    \begin{minipage}[t]{0.47\columnwidth}
        \centering
        \includegraphics[width=\columnwidth]{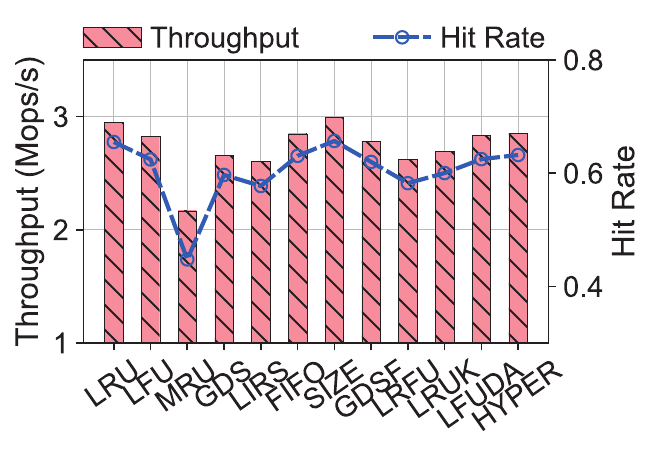}
        \caption{The throughput and hit rates of 12 algorithms.}
        \label{fig:policies-tpt-hr}
    \end{minipage}
\end{figure}

\noindent\textbf{Adapt to changing memory sizes.}
Figure~\ref{fig:mem-hr} shows the hit rate of \dmc when we dynamically increase the memory space.
The hit rate of \dmc approaches \dmc-LRU for most cases, outperforming \dmc-LFU.
When the cache size is 20\% and 30\% footprint size, the hit rate of \dmc-LFU exceeds \dmc-LRU.
\dmc performs better than both approaches because it can adaptively adjust its algorithm according to the affinity of caching algorithms on different cache sizes.


\subsection{Q4: Flexibility}
\noindent
To show that \dmc can flexibly integrate various caching algorithms, we integrate 12 commonly used caching algorithms into \dmc and evaluate their throughput, hit rate, and coding effort.
Since evaluating the feasibility of executing different caching algorithms is independent of workloads, we only show the throughput and hit rates on the \textit{webmail} workload in Figure~\ref{fig:policies-tpt-hr}.
Among all the algorithms, SIZE exhibits the best throughput and hit rate, while MRU exhibits the worst.
All these algorithms can be easily implemented in \dmc with less than 23 lines of code, as shown in Table~\ref{tab:code-info}.

\begin{figure}
    \begin{minipage}[t]{0.47\columnwidth}
        \centering
        \includegraphics[width=0.88\columnwidth]{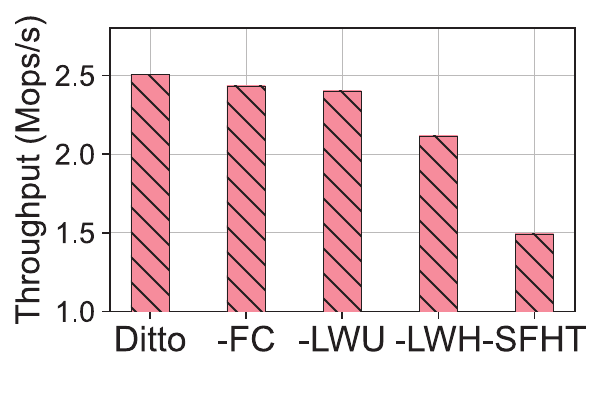}
        \caption{Contributions of different techniques on the \textit{webmail} workload.}
        \label{fig:design-tpt}
    \end{minipage}
    \hspace{2mm}
    \begin{minipage}[t]{0.47\columnwidth}
        \centering
        \includegraphics[width=0.98\columnwidth]{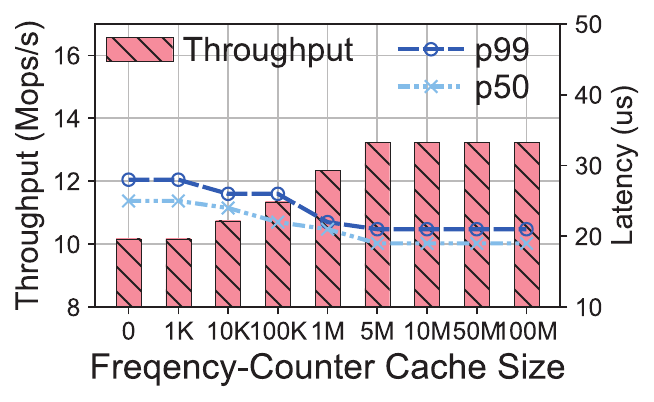}
        \caption{The YCSB-C performance of \dmc with different FC Cache sizes.}
        \label{fig:fc-tpt-lat}
    \end{minipage}
\end{figure}

\begin{table*}[tp]
    \small
    \centering
    \caption{\small LOCs and used access information of different caching algorithms on \dmc. $ts_{I}$ and $ts_{L}$ refer to the insert timestamp and the last access timestamp, respectively. S refers to the size of the object, F refers to the access frequency of the object, and M refers to the use of additional metadata. Details on the additional metadata M can be found in our open-source repository.}\label{tab:code-info}
    \begin{tabular}{l|c|c|c|c|c|c|c|c|c|c|c|c}
        \toprule[1pt]
        \textbf{Algs.} & LRU      & LFU & MRU      & GDS  & LIRS           & FIFO     & SIZE & GDSF    & LRFU        & LRUK & LFUDA & HYPERBOLIC     \\
        \midrule[0.5pt]
        \textbf{LOC}   & 9        & 9   & 9        & 14   & 12             & 9        & 9    & 14      & 17          & 23   & 14    & 11             \\
        \textbf{Info.} & $ts_{L}$ & F   & $ts_{L}$ & S & F, $ts_{L}$, M & $ts_{I}$ & S    & F, S & $ts_{L}$, M & M    & F, M  & $ts_{L}$, F, S \\
        \bottomrule[1pt]
    \end{tabular}
    \vspace{-0.1in}
\end{table*}

\subsection{Q5: Contribution of Each Technique}
\noindent
We show the contribution of techniques proposed in the paper by gradually disabling each technique of \dmc.
Due to the space limit, we show the performance of different techniques without miss penalties on the \textit{webmail} workload in Figure~\ref{fig:design-tpt}.
\dmc performs similarly on other workloads and more results can be found in our open-source repository.
The sample-friendly hash table (SFHT) improves the overall throughput by 42\% since it reduces the number of RDMA operations on data paths when updating the access information and sampling objects.
The lightweight history scheme (LWH) improves the throughput by 13\% due to the reduced number of RTTs when collecting regrets and maintaining eviction history.
Finally, the lazy weight update scheme (LWU) and the frequency-counter cache (FC) contribute to 4\% of the overall throughput because the reduced number of RDMA requests saves the message rate of the RNICs on MNs.

Figure~\ref{fig:fc-tpt-lat} shows the performance of \dmc under the YCSB-C benchmark with 256 clients and different FC cache sizes.
We limit FC cache size in MB since the size of each cache entry varies with the size of its recorded object ID.
We only show the result under YCSB-C due to the space limit. 
\dmc performs similarly on other workloads and more results can be found in our open-source repository.
The throughput increases from 10 Mops to 13.2 Mops with increased sizes of the FC cache since more \textsf{RDMA\_FAA}s can be cached locally to save the message rate of RNICs.
The tail latency drops from 28 us to 21 us due to the reduced number of RDMA operations and less contended network.
Also, the performance gain of the FC cache becomes insignificant when the size of the FC cache exceeds 5 MB, indicating that the FC cache can improve overall performance with small additional memory consumption on clients.

\section{Related Work}
\noindent
\textbf{Disaggregated Memory.}
Existing work on disaggregated memory can be classified into approaches that realize efficient memory disaggregation and approaches that design better applications.
The realization approaches use software-based~\cite{asplos23aguilera,osdi18shan,nsdi17gu,nsdi14dragojevic,osdi20ruan,eurosys18nitu,osdi20wang,eurosys20amaro}, hardware-based~\cite{isca09lim,genz,cxl}, and hybrid~\cite{sosp21lee,fast21wang,asplos22guo,arxiv21shan} techniques to efficiently achieve general-purpose disaggregated memory.
\dmc is orthogonal to these approaches since \dmc only assumes the underlying DM to be capable of executing \texttt{READ}, \texttt{WRITE}, \texttt{CAS}, and \texttt{FAA}.
Approaches that port applications on DM design important cloud applications, \eg key-value stores~\cite{atc20tsai,vldb22lee,fast23shen}, transactional storage systems~\cite{fast22zhang}, and data structures~\cite{sigmod22wang,atc21zuo,osdi23luo,fast23li,hotos19aguilera}, to achieve better resource efficiency.
The work most related to \dmc are memory-disaggregated key-value stores, \ie Clover~\cite{atc20tsai}, Dinomo~\cite{vldb22lee}, and FUSEE~\cite{fast23shen}.
However, all of them focus only on improving the performance for persistent and reliable data storage on DM, while \dmc is the first caching system that can efficiently execute various caching algorithms and adaptively select the best one on DM.

\textbf{In-Memory Caching Systems.}
Many approaches aim at improving the performance of Memcached~\cite{memcached} and Redis~\cite{redis}, the two most popular in-memory caching systems.
Some~\cite{hotcloud15cidon,nsdi16cidon,cloud14saemundsson} optimize the hit rate under objects of varying sizes.
Others~\cite{twemcache,nsdi13nishtala,nsdi21yang,nsdi13fan,nsdi14lim} improve memory efficiency and overall throughput.
The work closest to \dmc is CliqueMap~\cite{sigcomm21singhvi}, an RDMA-based caching system.
It uses one-sided \textsf{RDMA\_READ} for \textit{Get} operations and RPC for \textit{Set} operations, improving the throughput due to the higher bandwidth and CPU-bypass nature of one-sided \textsf{RDMA\_READ}.
However, all these approaches are designed and optimized for monolithic servers, which inevitably inherit the elasticity issues of monolithic servers.
\dmc exhibits better elasticity by leveraging the hardware benefits of DM.

\textbf{RDMA-Based KV stores.}
There are two types of RDMA-based KV stores, \ie server-centric and hybrid ones.
The former uses RDMA to construct fast RPC primitives and rely on server CPUs to access data~\cite{osdi16kalia,sigcomm14kalia,nsdi14dragojevic,nsdi14lim}.
The latter uses one-sided RDMA verbs to execute \textit{Get} operations and relies on server CPUs to execute \textit{Set} operations~\cite{osdi18wei,osdi20wei,atc13mitchell}.
Compared with these approaches, \dmc achieves efficient in-memory caching without relying on server-side CPUs.
Besides, the design of \dmc is not limited to RDMA. Other interconnects are also compatible.

\textbf{Caching Algorithms.}
Caching algorithms distinguish the hotness of objects using recency~\cite{hlru,value-aging}, frequency~\cite{tos17einziger} and other access information~\cite{atc17blankstein}, or combining various information together~\cite{nsdi18beckmann,atc17blankstein,gds,gdsf} to get higher hit rates.
Recently, there are many machine-learning-based adaptive caching algorithms~\cite {fast03megiddon,wdas02ari,fast21rodriguez,hotstorage18vietri}.
Among them, CACHEUS~\cite{fast21rodriguez} is the most related.
It uses regret minimization to adaptively select a better caching algorithm.
However, all these caching algorithms are designed for server-centric caching systems to optimize specific workloads.
\dmc, on the one hand, is designed for caching systems on DM where clients directly access data without involving CPUs in the memory pool.
On the other hand, \dmc is an adaptive caching framework where multiple caching algorithms can be integrated and adaptively selected according to workload and resource change.

\section{Conclusion}\label{sec:conclusion}
\noindent
We propose \dmc, the first caching system on the disaggregated memory architecture, to achieve better elasticity.
\dmc addresses the challenges of constructing a caching system on DM, \ie executing server-centric caching algorithms and dealing with inferior hit rates caused by dynamically changing resources and data access patterns.
A client-centric caching framework is proposed to efficiently execute caching algorithms on DM.
Various caching algorithms can be integrated with small coding efforts.
A distributed adaptive caching scheme is proposed to adapt to the resource and workload changes.
Experimental results show that \dmc effectively adapts to the resource and workload change on DM and outperforms the state-of-the-art caching system on monolithic servers by up to $9\times$ on YCSB synthetic workloads and $3.6\times$ on real-world key-value traces.
\begin{acks}
We sincerely thank our shepherd Marcos K. Aguilera and the anonymous reviewers for their constructive comments and suggestions.
This work is supported by \grantsponsor{rgc}{the Research Grants Council of the Hong Kong Special Administrative Region, China}{} (No. CUHK \grantnum{rgc}{14206921} of the General Research Fund), \grantsponsor{nsfc}{the National Natural Science Foundation of China}{} (Project No. \grantnum{nsfc}{62202511}), the \grantsponsor{nsfs}{Natural Science Foundation of Shanghai}{} (Project No. \grantnum{nsfs}{22ZR1407900}), and Huawei Cloud.
Pengfei Zuo is the corresponding author (pfzuo.cs@gmail.com).
\end{acks}

\bibliographystyle{plain}
\bibliography{main}

\end{document}